\documentclass[11pt,a4paper,preprintnumbers,nofootinbib]{revtex4}

\pdfoutput=1

\usepackage[T1]{fontenc}
\usepackage[utf8]{inputenc}
\usepackage{amsmath,amssymb,amsfonts,mathtools,bm}
\usepackage{hyperref}
\usepackage[final]{graphicx}
\usepackage{xcolor}
\usepackage{booktabs}
\usepackage{relsize}
\usepackage{lineno}
\usepackage{comment}

\usepackage{braket}

\def\ubar{\overline{u}}

\def\qbar{\overline{q}}
\def\lbar{\overline{\ell}}

\def\Babar{{\mbox{\slshape B\kern-0.1em{\smaller A}\kern-0.1em B\kern-0.1em{\smaller A\kern-0.2em R}}}}

\usepackage{cancel}

\usepackage{soul}

\usepackage[caption=false]{subfig}

\setstcolor{red}

\begin{document}

\title{Probing for New Physics  with Rare Charm Baryon $(\Lambda_c,\Xi_c,\Omega_c)$ Decays}

\author{Marcel Golz}
\email{marcel.golz@tu-dortmund.de}
\author{Gudrun Hiller}
\email{ghiller@physik.uni-dortmund.de}
\author{Tom Magorsch}
\email{tom.magorsch@tu-dortmund.de}
\affiliation{Fakult\"at f\"ur  Physik, TU Dortmund, Otto-Hahn-Str.4, D-44221 Dortmund, Germany}
\begin{abstract}
We analyze rare charm baryon decays within the standard model and beyond.
We identify all null test observables in unpolarized
$\Lambda_c \to p \ell^+ \ell^-$, $\ell=e, \mu$ decays, and  study the  new physics sensitivities.  We find that the longitudinal dilepton polarization fraction $F_L$ is sensitive to electromagnetic dipole couplings $C_{7}^{(\prime)}$,
and  to the right-handed  4-fermion vector coupling $C_{9}^\prime$. 
The forward-backward asymmetry, $A_{\rm FB}$, due to the GIM-suppression a standard model null test already, probes the  left-handed axialvector 4-fermion coupling $C_{10}$;
 its CP--asymmetry, $A_{\rm FB}^{\rm CP}$, probes CP--violating phases in  $C_{10}$.
Physics beyond the standard model  can induce branching ratios of dineutrino modes  $\Lambda_c \to p  \nu \bar \nu $ up to
a few times $10^{-5}$,  and one order of magnitude smaller if lepton flavor universality is assumed, while standard model rates are negligible.
Charged lepton flavor violation allows for striking signals into $e^\pm \mu^\mp$ final states, up to $10^{-6}$ branching ratios model-independently, 
and up to order $10^{-8}$ in leptoquark models.
Related three-body baryon decays $\Xi_c \to \Sigma  \ell \ell$, $\Xi_c \to \Lambda  \ell \ell$  and  $\Omega_c \to \Xi \ell \ell$  
offer similar opportunities to test the standard model with $|\Delta c|=|\Delta u|=1$ transitions.

\end{abstract}

\preprint{DO-TH 21/13}

\maketitle

\section{Introduction}

Rare $|\Delta c|=|\Delta u|=1$ processes are strongly Glashow-Iliopoulos-Maiani (GIM)-suppressed, a feature that blocks access to  standard model (SM) short-distance contributions
in simple observables such as branching ratios: either
because of overwhelming resonance effects (semileptonic  modes), or because branching ratios are too small (dineutrino modes) \cite{Burdman:2001tf}.
These very characteristics, on the other hand,  in addition to the approximate symmetries of the SM, give directions to form
clean null test  observables and to probe for a broad range of new physics (NP)  phenomena.

Search strategies in charm therefore necessarily differ from those in beauty, where the SM allows for precision studies, yet, a maturing experimental
program is also forcing here to optimize observables for a clean separation of NP from the SM background \cite{Cerri:2018ypt}.
The charm way beyond the SM also complements searches  from the strange and the beauty one, and offers unique insights into flavor physics from the up-sector,
{\it e.g.,}~\cite{Paul:2011ar,Fajfer:2015mia,deBoer:2015boa,Gisbert:2020vjx}.

In this work we study rare, three-body decays of charm baryons, extending previous works with charm mesons \cite{deBoer:2018buv,Bause:2019vpr}.
We study $\Lambda_c \to p \ell^+ \ell^-$, $\ell=e, \mu$ decays, and null test processes with charged lepton flavor violation  (cLFV) $\Lambda_c \to p e^\pm  \mu^\mp$,
and into dineutrinos  $\Lambda_c \to p \nu \bar \nu$. Hadronic transition form factors for $\Lambda_c \to p$ decays are available from lattice QCD \cite{Meinel:2017ggx}, and quark models \cite{Faustov:2018dkn}. We also present null test results for the related, spin $1/2 \to 1/2 + \ell \ell$ decays $\Xi_c \to \Sigma  \ell \ell$, $\Xi_c \to \Lambda  \ell \ell$ and  $\Omega_c \to \Xi \ell \ell$,  also induced by  $|\Delta c|=|\Delta u|=1$ quark transitions, with more strange  spectator quarks.  As for the $\Xi_c,\Omega_c$  decays  no form factor computations are available we employ the $\Lambda_c \to p$  ones together with $SU(3)_F$-flavor symmetries, if applicable.  While this can be improved in the future the impact of this approximation on the present study is sub-dominant
as we are mostly interested in null tests, and the SM-related uncertainties are dominated by resonance effects.

None of the rare semileptonic charmed baryon $(\Lambda_c,\Xi_c,\Omega_c)$ decays has been observed to date, but searches for
$\Lambda_c \to p \ell^+ \ell^-$, for both $\ell=e,\mu$ and  $\Lambda_c \to p e^\pm \mu^\mp $ by BaBar at the level of ${\cal{O}}(10^{-5})$ \cite{Lees:2011hb}, and $\Lambda_c \to p \mu^+ \mu^-$ by LHCb   at ${\cal{O}}(10^{-7})$ \cite{Aaij:2017nsd}  have been reported.
With  branching ratios at the order of $10^{-8}$ ($\ell^+ \ell^-$ final states), and as large as a few times $10^{-5}$ for dineutrinos \cite{Bause:2020xzj},
semileptonic rare charm baryon modes are suitable for experimental study at high luminosity flavor facilities, such as LHCb \cite{Cerri:2018ypt}, Belle II \cite{Kou:2018nap}, BES III \cite{Ablikim:2019hff}, and possible future machines \cite{Abada:2019lih}.

The plan of the paper is as follows: In Sec.~\ref{sec:theo} we discuss perturbative and non-perturbative contributions to exclusive rare charm baryon decay modes and present the fully differential distribution. Sec.~\ref{Sec:pheno_SM} contains the SM phenomenology. In Sec.~\ref{sec:SM_nulltests} we define null test observables based on the full angular distribution. In Sec.~\ref{sec:BSM} we provide model independent bounds on Wilson coefficients and highlight the NP sensitivity of the  null tests and their complementarity. In Sec.~\ref{sec:summary} we conclude.  In App.~\ref{app:WCs} we summarize the  perturbative SM contributions. The parametrization of hadronic matrix elements in terms of 
helicity form factors is given in App.~\ref{app:FF}. In App.~\ref{app:hel_amp} and \ref{app:lfv} we present the helicity amplitudes in terms of Wilson coefficients and form factors keeping finite lepton masses for the lepton flavor conserving and violating decay modes, respectively.

\section{$\Lambda_c \to p \ell^+ \ell^-$ Theory \label{sec:theo}}

We give general expressions for exclusive rare charm baryon decays in the SM and beyond. In Sec.~\ref{Sec:heff} we briefly review the perturbative SM contributions and introduce the weak Hamiltonian at the charm mass scale. Effects from intermediate resonances are discussed in Sec.~\ref{Sec:resonance_contributions}. We comment on charmed baryon decays other than $\Lambda_c\to p\ell^+ \ell^-$ in Sec.~\ref{Sec:otherbaryons}.
The fully differential  distribution for unpolarized $\Lambda_c \to p \ell^+ \ell^-$  decays is given in Sec.~\ref{Sec:angular-dist}. 

\subsection{An effective field theory approach to charm physics}\label{Sec:heff}

For the  phenomenological analysis of semileptonic $c\to u\ell^+\ell^-$ and $c \to u \bar{\nu}_i\nu_j$ processes we use the following effective Hamiltonian,
\begin{equation}
\mathcal{H}_{\rm eff} \supset -\frac{4G_F}{\sqrt2} \frac{\alpha_e}{4\pi} \biggl[ \sum_{k=7,9,10} \bigl( C_k O_k + C_k^\prime O_k^\prime \bigr) \,+\, \sum_{ij}\,\bigl(C_L^{ij}Q_L^{ij} +C_R^{ij}Q_R^{ij}\bigr)\biggr] \, , 
\label{eq:Heff}
\end{equation}
where the dimension six operators $O_k$ for dilepton and $Q_{L/R}^{ij}$ for dineutrino modes  are given as
\begin{equation}
\begin{split}
O_7 &= {m_c \over e} (\ubar_L \sigma_{\mu\nu} c_R) F^{\mu\nu} \,,\quad\quad \,O_7^\prime = {m_c \over e} (\ubar_R \sigma_{\mu\nu} c_L) F^{\mu\nu} \,, \\
O_9 &= (\ubar_L \gamma_\mu c_L) (\lbar \gamma^\mu \ell) \,, \quad\quad\quad \, O_9^\prime= (\ubar_R \gamma_\mu c_R) (\lbar \gamma^\mu \ell) \,, \\ 
O_{10} &= (\ubar_L \gamma_\mu c_L) (\lbar \gamma^\mu \gamma_5 \ell) \,,\quad\quad O_{10}^\prime= (\ubar_R \gamma_\mu c_R) (\lbar \gamma^\mu \gamma_5 \ell) \,, \\
Q_L^{ij} &=(\ubar_L \gamma_\mu c_L) (\bar{\nu}_{L\,j} \gamma^\mu \nu_{L\,i})\,, \quad Q_R^{ij} =(\ubar_R \gamma_\mu c_R) (\bar{\nu}_{L\,j} \gamma^\mu \nu_{L\,i})  \,,
\end{split}
\label{eq:operators}
\end{equation}
with the electromagnetic field strength tensor $F^{\mu\nu}$  and  $\sigma^{\mu\nu}=\frac{\text{i}}{2}\,[\gamma^\mu,\,\gamma^\nu]$. 
In the dineutrino operators  flavor indices $ij$ are made explicit, while we omit them for brevity in the charged dilepton operators $O_{9,10}^{(\prime)}$ unless necessary,
that is,  in the discussion of experimental constraints, lepton universality violation  and lepton flavor violation.
The primed operators $O_k^\prime$ are obtained from the $O_k$ by interchanging left-handed $(L)$ and right-handed $(R)$ chiral quark fields, $L\leftrightarrow R$. SM contributions  to the effective coefficients of $O_7$ and $O_9$ are generated by the charged-current four-quark operators 
\begin{equation}
P_1^q = (\ubar_L \gamma_\mu T^a\, q_L)(\qbar_L \gamma^\mu T^a\, c_L) \,, \quad 
P_2^q = (\ubar_L \gamma_\mu q_L)(\qbar_L \gamma^\mu c_L) \,, \quad  q=d,\,s,\, b \, , \\
\end{equation}
at the $W$ mass scale and induced by renormalization group running and mixing, as well as the two-step matching procedure, first at the $W$ mass scale and then at
 the $b$ mass scale~\cite{deBoer:thesis, deBoer:2017way, deBoer:2015boa}. Details are provided in Appendix~\ref{app:WCs}.
The SM contributions, including two-loop virtual corrections~\cite{deBoer:2017way}, depend on the dilepton invariant mass squared, $q^2$.
As shown in Fig.~\ref{fig:perturbative_sm_wcs} in App.~\ref{app:WCs} the non-vanishing SM contributions relevant to  $c \to u \ell^+ \ell^-$ are of the order of few permille for $|C_7^{\rm eff}|$ and few percent for $|C_9^{\rm eff}|$ above  $q^2>0.1\,\text{GeV}^2$.
Notably, the GIM-mechanism dictates $C_{10}^{\rm SM}=0$.
Contributions of primed operators are suppressed, and we neglect corresponding SM contributions. 

\subsection{Resonance contributions}\label{Sec:resonance_contributions}

The $\Lambda_c \to p \ell^+ \ell^-$ decay rate is dominated by intermediate $\bar q q$-resonances $M$ via $\Lambda_{c} \to p\,M$ and $M \to \ell^{+} \ell^{-}$. These contributions can be parameterized phenomenologically by Breit-Wigner distributions. For the spin-1 resonances $M = \rho(770),\,\omega(782),\,\phi(1020)$ this corresponds to an effective coefficient $C^{R}_{9}(q^{2})$ of the operator $O_9$ as
\begin{equation}
  \label{eq:resonances}
  C^{R}_{9}(q^{2}) = a_{\omega}e^{\text{i}\delta_{\omega}}\left(\frac{1}{q^{2} - m^{2}_{\omega} + \text{i}m_{\omega}\Gamma_{\omega}} - \frac{3}{q^{2} - m^{2}_{\rho} + \text{i}m_{\rho}\Gamma_{\rho}}\right)
        + \frac{a_{\phi}e^{\text{i}\delta_{\phi}}}{q^{2} - m^{2}_{\phi} + \text{i}m_{\phi}\Gamma_{\phi}}.
\end{equation}
Here, an isospin relation is employed to relate the $\rho$ and $\omega$ contributions to reduce the theoretical uncertainties. This approach is also necessary due to the lack of experimental information on $\mathcal{B}(\Lambda_c\to p\rho)$, and supported from charmed meson data~\cite{Bause:2019vpr}. The remaining two moduli of the resonance parameters $a_\omega$ and $a_\phi$ are fixed by experiment using
\begin{equation}
  \label{eq:resonance_parameters_extraction}
  \mathcal{B}_{C_9^R}(\Lambda_{c} \to  p\mu^{+}\mu^{-}) =  \mathcal{B}(\Lambda_{c} \to pM)\mathcal{B}(M \to \mu^{+}\mu^{-})\,,\quad\text{with }M=\omega,\,\phi\,.
\end{equation}
Taking the  respective branching fractions, masses and decay widths  from the PDG~\cite{Zyla:2020zbs} we obtain
\begin{equation}
\label{eq:a_fac}
\begin{split}
a_\omega=0.065\pm0.016\,,\\
a_\phi=0.110\pm0.008\,,
\end{split}
\end{equation}
in excellent agreement with Ref.~\cite{Meinel:2017ggx}.
As in semileptonic rare $b$-decays, a dispersion relation plus $e^+ e^- \to hadrons$-data \cite{Kruger:1996cv} can provide an alternative resonance  parameterization
to (\ref{eq:resonances}). It also
invokes factorization and fudge factors $a_M$, however,  avoids  double-counting by extracting the sum of long- and short-distance contributions experimentally, see 
also~\cite{Feldmann:2017izn} and \cite{Bharucha:2020eup} for application to $D\to\rho\ell\ell$ and $D\to\pi\ell\ell$ decays, respectively.
Since in charm, unlike in beauty, $C_9^R$ dominates over the SM short-distance contribution, see Fig.~\ref{fig:perturbative_sm_wcs},  and the parameterizations yield  
similar distributions in $D$-decays  within their sizable, notably, phase-induced uncertainties \cite{Bharucha:2020eup,Bause:2019vpr}, we employ the significantly simpler ansatz (\ref{eq:resonance_parameters_extraction}) for null test analyses in charmed baryons.

We also consider the contribution of  pseudoscalars  $\eta, \eta^\prime$ to  the  $\Lambda_{c} \to  p\mu^{+}\mu^{-}$ branching ratio.
While  leptonically decaying $\eta, \eta^\prime$ contribute to $c \to u \ell^+ \ell^-$ transitions, they are strongly localized \cite{Bause:2019vpr}, and outside of
target $q^2$-regions (\ref{eq:q2bins}). 
Neglecting lepton masses  we find,  in agreement with Ref.~\cite{Das:2018sms},
no interference terms between (axial-)vector contributions and those of the additional pseudoscalar operator $O_{P}= (\ubar_L c_R) (\lbar \gamma_5 \ell)$ in (\ref{eq:Heff}).
We  model the effects from the $\eta, \eta^\prime$-resonances via an effective coefficient
\begin{equation}
\label{eq:cp_res}
C^{R}_{P}(q^{2}) = a_{\eta}\left(\frac{1}{q^{2} - m^{2}_{\eta} + \text{i}m_{\eta}\Gamma_{\eta}} + \frac{e^{\text{i}\delta_{\eta^\prime}}}{q^{2} - m^{2}_{\eta^\prime} + \text{i}m_{\eta^\prime}\Gamma_{\eta^\prime}} \right) ,
\end{equation}
assuming the resonance parameter of the $\eta$ and the $\eta^\prime$ of similar size,
 $a_\eta\sim a_{\eta^\prime}$. 
Using masses, widths and branching fractions from the PDG~\cite{Zyla:2020zbs} and
\begin{equation}
  \mathcal{B}_{C_P^R}(\Lambda_{c} \to  p\mu^{+}\mu^{-}) =  \mathcal{B}(\Lambda_{c} \to p\eta)\mathcal{B}(\eta \to \mu^{+}\mu^{-})\, , 
\end{equation} 
we obtain $a_\eta \sim 5.8\cdot10^{-4}$, much smaller than the corresponding terms for vector mesons (\ref{eq:a_fac}).
The impact of $\eta, \eta^\prime$ on the differential and integrated $\Lambda_{c} \to  p\mu^{+}\mu^{-}$ branching ratio is discussed in Sec.~\ref{sec:BR}.

\subsection{Charmed baryons other than $\Lambda_c$}\label{Sec:otherbaryons}

 Three-body   spin $1/2 \to 1/2 + \ell \ell$, rare charm baryon decays other than   $\Lambda_c\to p \ell\ell$ can be studied in a similar manner, specifically $\Xi_c^0\to \Sigma^0 \ell\ell $, $\Xi_c^0 \to \Lambda^0  \ell \ell$, $\Xi_c^+\to \Sigma^+ \ell\ell $ and $\Omega_c^0\to\Xi^0 \ell\ell$ decays.
 Due to the lack of explicit form factor calculations, we use flavor symmetries and employ the form factors from $\Lambda_c\to p$, with obvious modifications in kinematic factors
 in (\ref{eq:ff1})-(\ref{eq:ff4}). Note that the $\bar u c$-type currents conserve $U$-spin and violate isospin by $|\Delta I|=1/2$.
 \footnote{This continues to hold for the whole  decay matrix elements of  the  effective Hamiltonian (\ref {eq:operators}), that is, ignoring four-quark operators.}
 Specifically,  $\Xi_c^+\to \Sigma^+$ is related by U-spin to $\Lambda_c\to p$, that is, the complete exchange of a $d$-quark with an $s$-quark.
 On the other hand, the $\Xi_c^0\to \Sigma^0$ form factors receive a factor $1/\sqrt{2}$ relative to the  $\Xi_c^+\to \Sigma^+$ ones
 due to isospin \cite{Adolph:2022ujd}. Then, U-spin relates $\Xi_c^0\to \Sigma^0$ form factors to $\Xi_c^0\to \Lambda^0$ ones \cite{Dery:2020lbc,Wang:2021uzi} with a relative factor of $\sqrt{3}$. For $\Omega_c^0$ decays  no relation with the $\Lambda_c$ exists since  the $\Omega_c^0$ sits in another $SU(3)_F$-multiplet \cite{Zyla:2020zbs}. 
 Due to the lack of other input we use the same form factors as for $\Lambda_c\to p$.
 To summarize, for any of the ten form factors commonly denoted here as $f_{A \to B}$, we employ
 $f_{\Lambda_c\to p}=f_{\Xi_c^+\to \Sigma^+}=\sqrt{2} f_{\Xi_c^0\to \Sigma^0}=\sqrt{6} f_{\Xi_c^0\to \Lambda^0} \simeq  f_{\Omega_c^0\to\Xi^0}$  
 \cite{Adolph:2022ujd}.
 
None of the branching ratios $\mathcal{B}(\Xi_c\to \Sigma (\Lambda) M)$ and $\mathcal{B}(\Omega_c\to\Xi M)$ with $M=\rho,\,\omega,\,\phi$ has been observed so far.
To estimate the $a_M$-resonance parameters, we 
 relate the $\Xi_c, \Omega_c$-branching ratios with two-body phase space factors,  lifetimes and the aforementioned flavor  factors to the ones of $\Lambda_c^+\to p$ modes. 
 One obtains the $a_{\omega},\,a_{\phi}$ factors  given in Table~\ref{tab:afacs}, which are similar to Eq.~\eqref{eq:a_fac}, as expected. 
 This ansatz also gives $\mathcal{B}(\Xi_c^0\to \Lambda^0 \phi)\sim1.2\cdot10^{-4}$, consistent with but on the lower end of the first evidence measurement $\mathcal{B}(\Xi_c^0\to \Lambda^0 \phi)=(4.9\pm1.5)\cdot10^{-4}$~\cite{Zyla:2020zbs}, which suggests $a_\phi\sim0.2$.

 \begin{table}[!t]
 \centering
  \caption{$a_{\omega},\,a_{\phi}$ parameters for various rare charm baryon modes,
   see  text for details.}
 \label{tab:afacs}
 \begin{tabular}{l||c|c|c|c|c}
& $\Lambda_c\to p$ & $\Xi_c^+\to\Sigma^+$ & $\Xi_c^0\to\Sigma^0$ & $\Xi_c^0\to\Lambda^0$ & $\Omega_c^0\to\Xi^0$ \\
\hline 
$a_\omega$ & $0.065 \pm 0.016$ & $\sim 0.06$ & $\sim 0.06$ & $\sim 0.06$& $\sim 0.05$\\
$a_\phi$ & $0.110 \pm 0.008$ & $\sim0.1$ & $\sim 0.1$ & $\sim 0.1$ & $\sim 0.09$ \\
 \end{tabular}
\end{table}

\subsection{Fully differential distribution  for $m_\ell \neq 0$ \label{Sec:angular-dist}}

For the calculation of SM and BSM contributions to the two-fold differential decay rate of $\Lambda_c \to p \ell^+ \ell^-$ we utilize the helicity formalism~\cite{Haber:1994pe, Gratrex:2015hna}. Details on the calculation of the relevant helicity amplitudes are collected in App.~\ref{app:hel_amp}.
The fully differential angular distribution for unpolarized $\Lambda_c$ can be written as
\begin{equation}
  \frac{\text{d}^2\Gamma}{\text{d}q^2\text{d}\cos\theta_\ell}=\frac{3}{2}\,(K_{1ss}\,\sin^2\theta_\ell\,+\,K_{1cc}\,\cos^2\theta_\ell\,+\,K_{1c}\,\cos\theta_\ell)\,.
  \label{eq:angl_distr}
\end{equation}
Here,  $\theta_\ell$ is the angle of the $\ell^+$ with respect to the negative direction of flight of the $\Lambda_c$  in the dilepton rest frame.
The $q^2$-dependent coefficents are given as \cite{Gutsche:2013pp} 
\begin{align}
  \label{eq:2}
  \begin{split}
    &K_{1ss} = q^{2}v^{2}\left(\frac{1}{2}U^{11+22} + L^{11+22}\right) + 4m^{2}_{\ell}\left(U^{11}+L^{11}+S^{22}\right),\\
    &K_{1cc} = q^{2}v^{2}U^{11+22} + 4m^{2}_{\ell}\left(U^{11}+L^{11}+S^{22}\right),\\
    &K_{1c} = -2q^{2}vP^{12},
  \end{split}
\end{align}
in agreement with our own computation and  \cite{Blake:2017une}. Here,
$v = \sqrt{1-\frac{4m^{2}_{\ell}}{q^{2}}}$, $U^{11+22}=U^{11} + U^{22}$ and $L^{11+22}=L^{11} + L^{22}$. The $q^2$-dependent  terms $U,\, L,\, S,\, P$ denote quadratic expressions of helicity amplitudes and correspond to unpolarized transverse, longitudinal, scalar and parity violating contributions, respectively. For further details we refer to App.~\ref{app:hel_amp}. Here we give the respective expressions in terms of Wilson coefficients and hadronic form factors, $f_i(q^2), g_i(q^2), i=+,\perp,0$ and $h_j(q^2), \tilde h_j(q^2), j=+, \perp$, defined in  App.~\ref{app:FF}.  For brevity we omit their $q^2$-dependence in the following:
\begin{align}
  \begin{split}
  \label{eq:36}
  U^{11} =\phantom{-} 4N^{2}\cdot\bigg[\,& \bigg|(C_{7} + C_7^\prime)\,\frac{2m_{c}}{q^{2}}(m_{\Lambda_{c}} + m_{p})\,h_{\perp} + (C_{9} + C_9^\prime)\,\,f_{\perp}\bigg|^{2}\cdot s_{-}\\
  + &\bigg|(C_{7} - C_7^\prime)\,\frac{2m_{c}}{q^{2}}(m_{\Lambda_{c}} - m_{p})\,\tilde{h}_{\perp} + (C_{9} - C_9^\prime)\,\,g_{\perp}\bigg|^{2}\cdot s_{+} \, \bigg]\,,\\
  L^{11} =\phantom{-} \frac{2N^{2}}{q^2}\cdot\bigg[\,&\bigg|(C_{7} + C_7^\prime)\,2m_{c}\,h_{+} + (C_{9} + C_9^\prime)\,(m_{\Lambda_{c}} + m_{p})\,f_{+}\bigg|^{2}\cdot s_{-}\\
  + &\bigg|(C_{7} - C_7^\prime)\,2m_{c}\,\tilde{h}_{+} + (C_{9} - C_9^\prime)\,(m_{\Lambda_{c}} -  m_{p})\,g_{+}\bigg|^{2}\cdot s_{+} \, \bigg]\,,\\
  U^{22} =\phantom{-} 4N^{2}\cdot\bigg[\,&\bigg|(C_{10} + C_{10}^\prime)\, f_{\perp}\bigg|^{2}\cdot s_{-} +\,\, \bigg|(C_{10} - C_{10}^\prime)\, g_{\perp}\bigg|^{2}\cdot s_{+}\, \bigg]\,,\\
  L^{22} =\phantom{-} \frac{2N^{2}}{q^2}\cdot\bigg[\,&\bigg|(C_{10} + C_{10}^\prime)\, (m_{\Lambda_{c}} + m_{p})\,f_{+}\bigg|^{2}\cdot s_{-} +\,\, \bigg|(C_{10} - C_{10}^\prime)\, (m_{\Lambda_{c}} - m_{p})\,g_{+}\bigg|^{2}\cdot s_{+}\, \bigg]\,,\\
  S^{22} = \phantom{-}\frac{2N^{2}}{q^2}\cdot\bigg[\, &\bigg|(C_{10} + C_{10}^\prime)\,(m_{\Lambda_{c}} - m_{p})\,f_{0}\bigg|^{2}\cdot s_{+} +\,\, \bigg|(C_{10} - C_{10}^\prime)\,(m_{\Lambda_{c}} + m_{p})\,g_{0}\bigg|^{2}\cdot s_{-}\, \bigg]\,,\\
  P^{12} = -8N^{2}\cdot\bigg[\,&\text{Re}\big((C_{7} - C_7^\prime)\,(C^{*}_{10} + C^{\prime *}_{10})\big)\,\frac{m_{c}}{q^{2}}(m_{\Lambda_{c}} - m_{p})\,f_{\perp}\,\tilde{h}_{\perp}\\
  +\,&\text{Re}\big((C_{7} + C_7^\prime)\,(C^{*}_{10} - C^{\prime *}_{10})\big)\,\frac{m_{c}}{q^{2}}(m_{\Lambda_{c}} + m_{p})\,g_{\perp}\,h_{\perp} \\+ &\,\text{Re}\big(C_{9} C^{*}_{10} - C_9^\prime C^{\prime *}_{10}\big)\, g_{\perp}\,f_{\perp}\bigg]\cdot\sqrt{s_{+}s_{-}}\,.
  \end{split}
\end{align}
The factor $N^2$ is a global normalization and is given by  $N^2={\frac{G^{2}_{F}\alpha^{2}_{e}v\sqrt{\lambda(m^{2}_{\Lambda_{c}},\,m^{2}_{p},\,q^{2})}}{3\cdot 2^{11}\pi^{5}m^{3}_{\Lambda_{c}}}}$ with the Källén function $\lambda(a,\,b,\,c)=a^2+b^2+c^2-2\,(ab+ac+bc)$ and $s_{\pm} = (m_{\Lambda_c} \pm m_p)^2 - q^2$.

Note, in the SM all terms vanish except for $U^{11}$ and $L^{11}$.
Therefore, $K_{1c}^{\rm SM} =0$. At the kinematic endpoint of zero hadronic recoil vanishes $s_-$ and $P^{12}/N^2=0$ also in the presence of NP.
At both zero $q^2=(m_{\Lambda_{c}} - m_{p})^2$ and maximal $q^2=4 m_\ell^2$ recoil endpoints further holds $K_{1ss}=K_{1cc}$ model-independently.

The observables discussed in Sec.~\ref{sec:SM_nulltests} can be expressed in terms of $K_{1ss},\,K_{1cc},\,K_{1c}$ and their CP--conjugated counterparts $\bar K_{1ss},\,\bar K_{1cc},\,\bar K_{1c}$. For instance, the fractions of longitudinal and transverse polarization of the dilepton system are, respectively, given by
              \begin{equation}
              F_L=\frac{2\,K_{1ss}-K_{1cc}}{2\,K_{1ss}+K_{1cc}}\,, \qquad
              F_T=1-F_L=\frac{2\,K_{1cc}}{2\,K_{1ss}+K_{1cc}}\,.
              \label{eq:fl}
              \end{equation}
The $q^2$-differential decay rate  is obtained by integrating (\ref{eq:angl_distr}) over the angle $\theta_\ell$ within $-1 \leq \cos\theta_\ell \leq +1$
\begin{equation}
\frac{\text{d}\Gamma}{\text{d}q^2}=\int_{-1}^{1}\frac{\text{d}^2\Gamma}{\text{d}q^2\text{d} \! \cos\theta_\ell}\,\text{d} \! \cos\theta_\ell=2\,K_{1ss}+K_{1cc}\, .
\label{eq:diffbrratio}
\end{equation}
The $\Lambda_c \to p \ell^+ \ell^-$ decay rate follows as $\Gamma=\int_{q^2_{\text{min}}}^{q^2_{\text{max}}}\,(2\,K_{1ss}+K_{1cc})\,\text{d}q^2$,
where the phase space is limited by $q^2_{\text{min}}=4 m_\ell^2$ and $q^2_{\text{max}}=(m_{\Lambda_c}-m_p)^2$,  but we will also employ other $q^2$-cuts to enhance the NP sensitivity.
Specifically, we introduce, targeting muons,
\begin{equation}\label{eq:q2bins}
\begin{split}
\text{full }q^2\text{ range:}&\quad\quad 2\,m_\mu \leq \sqrt{q^2} \leq m_{\Lambda_c}-m_p\,,\\
\text{high }q^2\text{ range:}&\quad 1.25\,\text{GeV} \leq \sqrt{q^2} \leq m_{\Lambda_c}-m_p\,,\\ 
\text{low }q^2\text{ range:}&\quad \quad 2\,m_\mu \leq \sqrt{q^2} \leq 0.525\,\text{GeV}\,.
\end{split} 
\end{equation}
Here, the low $q^2$ region corresponds to the region below the $\eta$, whereas the high $q^2$ region
is above the $\phi$.

We discuss $F_L$, $\text{d} {\cal{B}}/\text{d}q^2=\tau \text{d} \Gamma/\text{d}q^2$, with $\Lambda_c$-lifetime $\tau$, and the resulting branching ratios in the next section \ref{Sec:pheno_SM}.

\section{SM Phenomenology \label{Sec:pheno_SM}}

We discuss the SM prediction for the $q^2$-differential and integrated branching ratios in Section \ref{sec:BR}, and $F_L$ in Section \ref{sec:fl}.

\subsection{Branching ratio  \label{sec:BR}}

\begin{figure}[!t]\centering
\includegraphics[width=0.7\textwidth]{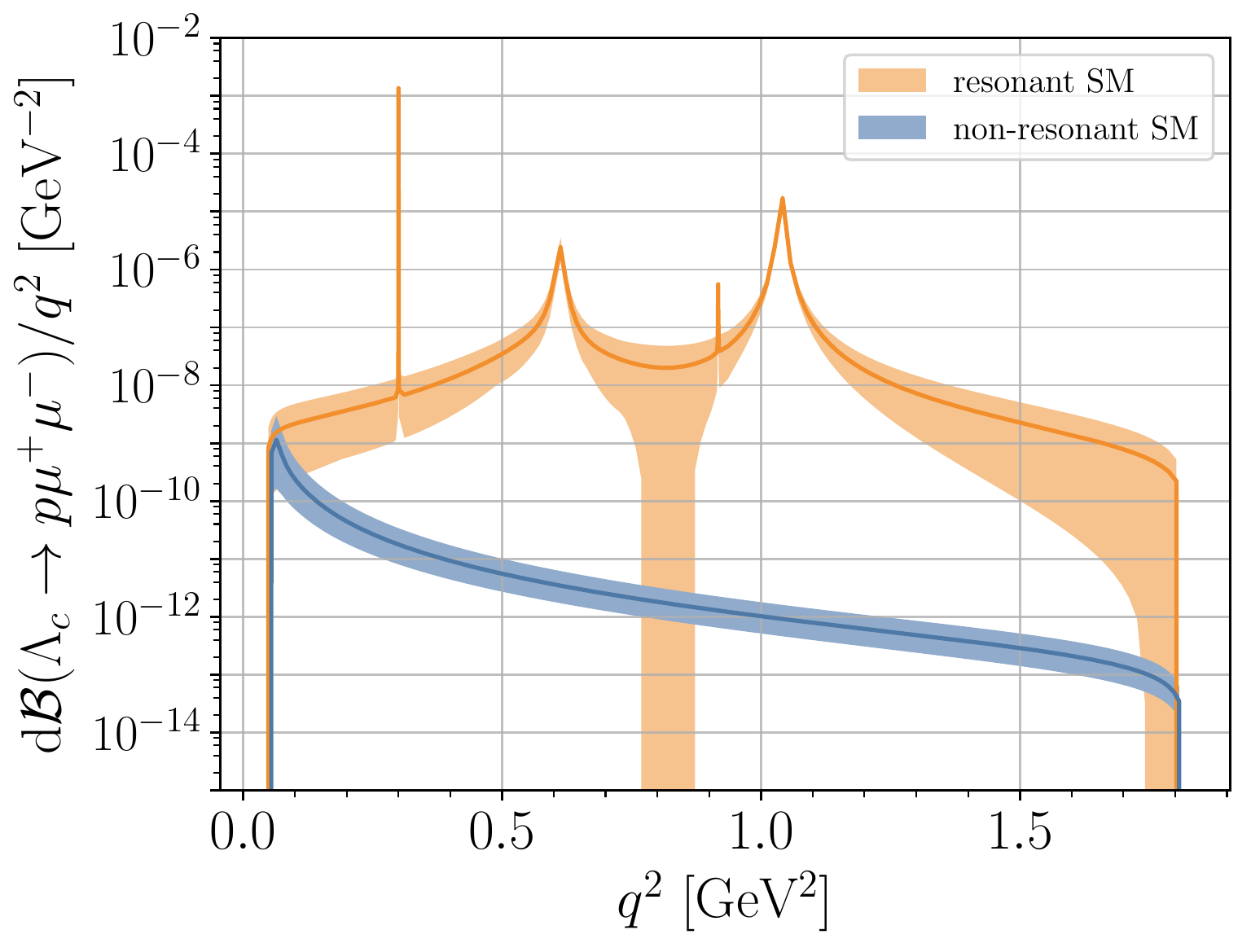}
\caption{The differential SM branching fraction of $\Lambda_c\to p \mu^+\mu^-$. In blue (orange) the short-distance (pure resonant) contributions are shown. Their widths indicate theoretical uncertainties of hadronic form factors, resonance parameters and scale uncertainty $\mu_c$. The central line in the resonant case is obtained by averaging after varying strong phases.}
\label{fig:SMBR}
\end{figure}

The differential branching ratio of $\Lambda_c\to p \mu^+\mu^-$ decays  is shown in Fig.~\ref{fig:SMBR} in blue and orange for the non-resonant and resonant contributions, respectively. Uncertainties arise from the form factors, and for the non-resonant contribution mainly due to the scale uncertainty of the Wilson coefficients.
 The main source of uncertainty for the resonant contributions  are the strong phases $\delta_\omega$ and $\delta_\phi$ in Eq.~\eqref{eq:resonances}. The central value curve of the resonance band is obtained by averaging after varying both phases from $-\pi$ to $\pi$.   Clearly, SM contributions from the effective coefficients $C_{7,\,9}^{\rm eff}$, see Fig.~\ref{fig:perturbative_sm_wcs}, are negligible in the branching ratio in the full $q^2$ range. Therefore, we only consider resonant contributions for the SM contributions in the subsequent analysis, unless otherwise stated.

In Fig.~\ref{fig:SMBR} additional 
contributions of  $\eta,\,\eta^\prime$-resonances decaying leptonically  to the differential branching fraction are displayed using (\ref{eq:cp_res}) and \cite{Das:2018sms}
\begin{equation}
 \text{d}\mathcal{B}/\text{d}q^2 \vert_{\eta,\,\eta^\prime}=6 \tau q^2 N^2\,\left(g_0^2\frac{(m_{\Lambda_c}+m_p)^2}{m_c^2}\,s_-+f_0^2\frac{(m_{\Lambda_c}-m_p)^2}{m_c^2}\,s_+\right)\,\vert C_P^R \vert^2\,.
\end{equation}
Although the $\eta$-peak reaches values as large as $\mathcal{O}(10^{-3})$, its contribution to $\mathcal{B}(\Lambda_c\to p\mu^+\mu^-)$ is very small, $\mathcal{O}(10^{-9})$, 
due to its narrow width, and negligible in the branching ratio. Similar results are obtained for the $\eta^\prime$. In the following we therefore do not consider the $\eta,\,\eta^\prime$-contributions any further, as they are
quite located and outside the primary $q^2$-bins of interest (\ref{eq:q2bins}).

The LHCb collaboration has
obtained the $90\,\%$ CL upper limit~\cite{Aaij:2017nsd}
\begin{equation}\label{eq:lhcb_upper_limit}
\mathcal{B}_{\text{LHCb}}(\Lambda_c\to p \mu^+\mu^-) < 7.7\,\times\,10^{-8}\,.
\end{equation}
Bounds on the electron modes are two orders of magnitude weaker, 
$\mathcal{B}(\Lambda_c\to p e^+e^-) < 5.5\,\times\,10^{-6}$ at $90\,\%$ CL (BaBar) \cite{Lees:2011hb}.
The upper limit in Eq.~\eqref{eq:lhcb_upper_limit} is based on a search for events outside the $\omega$ and $\phi$ windows -- $\pm40\,$MeV around the resonance masses --
and subsequent use of  a phase space model to "fill the gaps". As neither  the latter  nor the corresponding correction factor to the event rate is  publicly available 
a  straightforward comparison with theory predictions is not possible. We therefore suggest to avoid extrapolations in future experimental analyses and instead quote
plain rates with the cuts they are based on.

Using Eqs.~\eqref{eq:resonances} and~\eqref{eq:a_fac}  we compute the  $\Lambda_c\to p \mu^+\mu^-$ branching ratio in the full $q^2$-region minus the $\pm40\,$MeV windows around the $\omega$ and $\phi$ resonances
and obtain
\begin{equation}\label{eq:smbrlambda}
\mathcal{B}_{\text{SM}}(\Lambda_c\to p \mu^+\mu^-) = (1.9^{+1.8}_{-1.5})\,\times\,10^{-8}\,.
\end{equation}
For the decays other than the $\Lambda_c$ we employ the resonance parameters in Table~\ref{tab:afacs}.
The branching ratios  for $\Xi_c^+\to \Sigma^+\mu^+\mu^-$ ($\Omega_c^0\to\Xi^0\mu^+\mu^-$) are enhanced with respect to Eq.~\eqref{eq:smbrlambda} by factors of $\sim 1.8\,(1.3)$, which is mainly driven by the different lifetimes. For the same reason and the suppression factors in the form factors the branching ratios of $\Xi_c^0\to\Sigma^0 (\Lambda^0)\mu^+\mu^-$ are reduced with respect to Eq.~\eqref{eq:smbrlambda} by a factor $\sim 0.4\,(0.2)$. In the remainder of this work we analyse null tests of the SM and present our results in terms of contributions to the decay $\Lambda_c\to p \ell\ell$. However, all observables can also be studied for the other modes.
We recall that dominant decay modes are  $\Sigma^+ \to p \pi^0, n \pi^+$, $\Sigma^0 \to \Lambda \gamma$, $\Lambda^0\to p \pi^-, n \pi^0$ and $\Xi^0 \to \Lambda \pi^0$~\cite{Zyla:2020zbs}.

\subsection{$F_L$ \label{sec:fl}}

\begin{figure}[!ht]
        \includegraphics[width=0.49\textwidth]{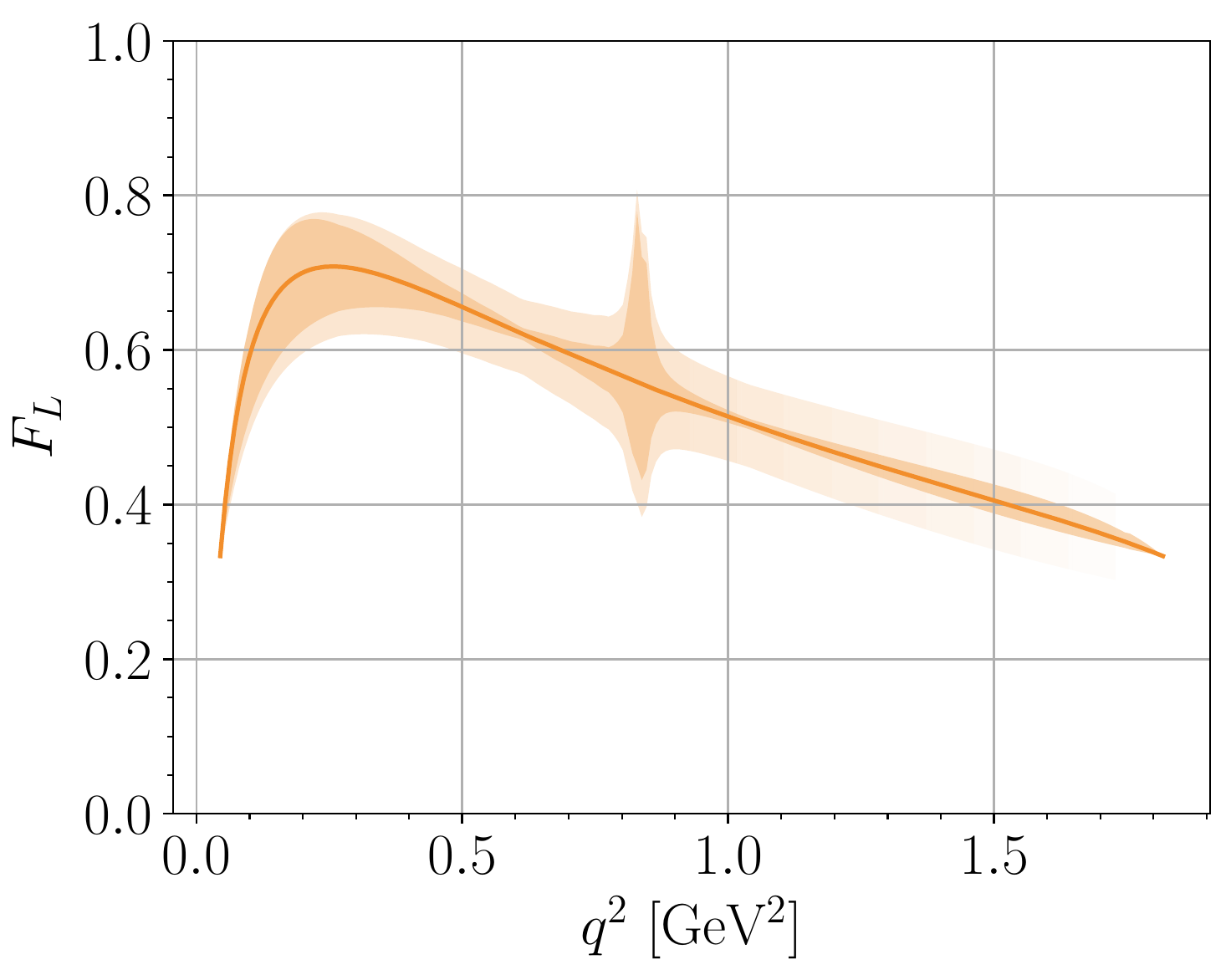}
     \caption{The fraction of longitudinally polarized dimuons $F_L$ (\ref{eq:fl}) in the SM.
     The solid dark line corresponds to the contribution from $C_R^9$ only. The orange band includes in addition 
     $C_{7,9}^{\rm eff}(q^2)$. The pale bands illustrate a $10 \%$ splitting between the longitudinal and the transverse fudge factors, which is forbidden at zero recoil, see text.
     At both zero and maximum recoil holds $F_L=1/3$. }
     \label{fig:flsm}
 \end{figure}

The fraction of longitudinally polarized dileptons, $F_L$, defined in  (\ref{eq:fl}), is mostly dictated by helicity. 
As is apparent from the expressions given in Sec.~\ref{Sec:angular-dist},
at the zero recoil endpoint, holds model-independently $F_L(s_-=0)=1/3$, as in $B \to K^* \ell^+ \ell^-$, as this kinematic point has higher symmetry \cite{Hiller:2013cza}, also observed for baryons \cite{HillerZwicky21}.
In particular, the four-momenta of the decaying hadron, the final hadron, and the dilepton system, $q^\mu$, are proportional to each other at zero recoil, allowing 
for fewer Lorentz structures, hence relations between helicity amplitudes.
At the maximum  hadronic recoil endpoint, we also obtain $F_L(4 m_\ell^2)=1/3$, in agreement with
\cite{Faustov:2018dkn}~\footnote{This is in conflict with Fig.~7 of \cite{Meinel:2017ggx}.}.
This is no coincidence, as at maximum recoil the four-momenta of the leptons are the same, causing  also here the angular distribution (\ref{eq:angl_distr}) to become flat,
$d \Gamma/d \cos \theta_\ell=const$.
In between these kinematic endpoint constraints $F_L$ rises due to the dominant longitudinal contributions with $1/q^2$-enhancement.
This can be seen in Fig.~\ref{fig:flsm}, where $F_L$ is shown  in the SM.
The structures around $q^2\sim 0.8 \, {\mbox{GeV}}^2$ are caused by interference of $C^R_9$, which has significant uncertainties in the region between the $\omega$ and the 
$\phi$, see   Fig.~\ref{fig:SMBR}, with the perturbative effective
dipole coefficient $C_{7}^{\rm eff}$. Due to the dominance of $C_9^R$, numerically the impact of   $C_{9}^{\rm eff}$ is negligible.
The shape of $F_L$ can change dramatically with BSM physics, however,
 its behavior can still be understood by helicity arguments, see Sec.~\ref{sec:BSM_nulltests} for details.

As we defined $F_L$ with normalization to the differential decay rate (\ref{eq:fl}), rather than the integrated rate, the cancellation of resonance effects between numerator and denominator is highly efficient.
Note, this  cancellation  hinges on  the fudge factors $a_M$ being  polarization-independent. In principle, polarization-specific fudge factors  could be taken into account if 
$F_L$ is measured
in $\Lambda_c \to p M$ decays. We simulate effects of the order $10 \%$ (pale bands). We stress, however, that endpoint constraints prohibit polarization-dependent fudge factors 
at zero recoil. We therefore expect the SM prediction from the phenomenological parameterization to be most robust towards maximum $q^2$.

We  provide an integrated polarization fraction in the SM,  integrating full and high $q^2$ regions, as in (\ref{eq:q2bins}), respectively,
\begin{equation}\label{eq:smfl2}
F_L|_{\text{SM}}(\Lambda_c\to p \mu^+\mu^-) = \int_{q^2_{\text{min}}}^{q^2_{\text{max}}}\,\frac{2\,K_{1ss}-K_{1cc}}{2\,K_{1ss}+K_{1cc}} \,\text{d}q^2 \simeq 0.94 \pm 0.05 \, ,
\end{equation}
\begin{equation}\label{eq:smfl3}
F_L|_{\text{SM}}(\Lambda_c\to p \mu^+\mu^-) = \int_{1.25^2 \, \text{GeV}^2}^{q^2_{\text{max}}}\,\frac{2\,K_{1ss}-K_{1cc}}{2\,K_{1ss}+K_{1cc}} \,\text{d}q^2 \simeq 0.093 \pm 0.004 \, . 
\end{equation}
Normalizing to the integrated decay rate, with cuts corresponding to (\ref{eq:smbrlambda}),  cutting $\pm40\,$MeV around the  $\omega,\phi$ masses, we obtain instead $0 \leq \tilde F_L \leq 1$.
As expected, here uncertainties do not cancel and remain sizable, and cover the full physical range.

\section{Null test observables}\label{sec:SM_nulltests}

We give the complete set of observables in three-body rare charm baryon decays  and identify null tests of the SM. All observables are based on the angular coefficients $K_{1ss},\,K_{1cc}$ and $K_{1c}$ (and their CP--conjugates), which are given in terms of Wilson coefficients and form factors in Eq.~\eqref{eq:2} and Eq.~\eqref{eq:36} and further details in App.~\ref{app:hel_amp}.

As already discussed in Sec.~\ref{Sec:pheno_SM} and illustrated in Fig.~\ref{fig:SMBR} the $q^2$-differential decay rate (\ref{eq:diffbrratio})
  has little sensitivity to BSM physics, due to resonance contributions (orange band), however, it can be used to form null tests, such as ratios of dimuon to dielectron rates, or CP--asymmetries, discussed below.
  An observable which is also not a null test  yet sensitive to NP is the fraction of longitudinal polarization of the dilepton system, $F_L$, defined in (\ref{eq:fl}).

In this work we consider in addition the following null test observables:

\begin{enumerate}
    \item The forward-backward asymmetry of the leptonic scattering angle,    is defined as
              \begin{equation}
              A_{\text{FB}}=\frac{1}{\text{d}\Gamma /\text{d}q^2}\,\left[\int_0^1\,-\,\int_{-1}^0\right]\frac{\text{d}^2\Gamma}{\text{d}q^2\text{d}\cos\theta_\ell}\,\text{d}\cos \theta_\ell = \frac{3}{2}\,\frac{K_{1c}}{2\,K_{1ss}+K_{1cc}}\,.
              \label{eq:afb}
              \end{equation}
              The angular observable $K_{1c}$ is proportional to linear combinations of $C_{10}$ and $C'_{10}$ and thus $A_{\text{FB}}$ constitutes a 
               clean, charm-specific null test of the SM.
              Note that at higher order in the fine structure constant  loop-induced contribution to the matrix element of $O_{10}$ arise,  
              at the level of $ \lesssim 10^{-4}$~\cite{deBoer:thesis,deBoer:2018buv}, that are entirely negligible for the present and foreseeable uncertainties.

    \item Lepton universality can be probed analogously to the ratios of charmed meson decays~\cite{Fajfer:2015mia, deBoer:2018buv, Bause:2019vpr} with the ratio
    \begin{equation}
    R^{\Lambda_c}_p={\int_{q^2_{\text{min}}}^{q^2_{\text{max}}} \frac{\text{d}\mathcal{B}(\Lambda_c\to p \mu^+\mu^-) }{\text{d}q^2} \text{d}q^2 }{\Big/}{\int_{q^2_{\text{min}}}^{q^2_{\text{max}}}  \frac{\text{d}\mathcal{B}(\Lambda_c\to p e^+e^-) }{\text{d}q^2} \text{d}q^2 }\,,\label{eq:R_ratio}
    \end{equation}
    where $q^2_{\text{min}}$ and $q^2_{\text{max}}$ can be chosen in bins or the full kinematically allowed region, but necessarily need to be equal for the muon and the electron mode. Due to the universality of the SM interactions $R^{\Lambda_c}_p$ is expected to be close to $1$ in the SM, such that $R^{\Lambda_c}_p-1$ is an excellent null test.

    \item The CP--asymmetry in the decay rate is defined as
              \begin{equation}
              A_{\text{CP}}=\frac{\text{d}\Gamma / \text{d} q^2 \,-\,\text{d}\bar\Gamma / \text{d} q^2  }{\text{d}\Gamma / \text{d} q^2 \,+\,\text{d}\bar\Gamma / \text{d} q^2 }=\frac{2\,K_{1ss}+K_{1cc} - 2\,\bar{K}_{1ss}-\bar{K}_{1cc}}{2\,K_{1ss}+K_{1cc} + 2\,\bar{K}_{1ss}+\bar{K}_{1cc}}\,,
              \label{eq:acp}
              \end{equation}
              where the bar indicates CP--conjugation. Since CP--violating effects in charm are strongly CKM-suppressed in the SM, 
              parametrically by $V_{cb}^*V_{ub}/(V_{cs}^* V_{us})$,
              $A_\text{CP}$ constitutes a null test of the SM which is sensitive to NP with a CP--violating phase.
      \item It is also possible to study the CP--asymmetry of an angular observable. Here, we study the CP--asymmetry of the forward-backward asymmetry, defined as
        \begin{equation}
            A_{\text{FB}}^{\text{CP}}=\frac{A_{\text{FB}}+\bar{A}_{\text{FB}}}{A_{\text{FB}}-\bar{A}_{\text{FB}}}= \frac{K_{1c}-\bar K_{1c}}{K_{1c}+\bar K_{1c}}\,,
        \label{eq:afbcp}
        \end{equation}
        studied previously in other decay modes \cite{Buchalla:2000sk,Paul:2011ar}. In the SM it is protected by GIM and small CP--phases.
        In the CP--conserving limit holds $A_{\text{FB}}=-\bar A_{\text{FB}}$, that is, $A_{\rm FB}$ is a CP-odd observable.
        The reason for this is that we keep for both  $\Lambda_c^+$ and $\Lambda_c^-$ decays  the lepton angle $\theta_\ell$  defined with respect to the same, positively charged lepton, while 
         CP--conjugation would exchange $\ell^+\leftrightarrow\ell^-$.  The CP--conjugated angular distribution for $\Lambda_c^-$ is thus obtained by exchanging $\theta_\ell \leftrightarrow \pi-\theta_\ell$,
         which, using  $\cos{(\pi-\theta_\ell)}=-\cos{\theta_\ell}$ results in an extra minus sign in (\ref{eq:angl_distr}) in front of $K_{1c}$.
    \item Charged lepton flavor violation: $\mathcal{B}(\Lambda_c\to p\ell\ell^\prime)$ with $\ell\neq\ell^\prime$.   Due to the negligibe SM contribution, suppressed by tiny neutrino masses,  any signal is NP.
    \item Dineutrino final states: $\mathcal{B}(\Lambda_c\to p\nu\bar\nu) =\sum_{i,j}\mathcal{B}(\Lambda_c\to p\nu_i\bar\nu_j)$\,. 
    Due to the negligibe SM contribution, suppressed by the GIM mechanism,  any signal is NP. In addition, branching ratios can reach upper limits depending on charged lepton flavor. This allows to test lepton universality and probe for cLFV \cite{Bause:2020xzj, Bause:2020auq}.
    
\end{enumerate}

We comment  on different conventions to normalize  $A_{\text{FB}}$ and $A_{\text{CP}}$ used in the literature. While in Eqs.~\eqref{eq:afb} and \eqref{eq:acp}, and also Eq.~\eqref{eq:fl}, the normalization
is to the $q^2$-differential decay rate, one may consider normalizing to the  rate $\Gamma$ integrated over different bins $[q^2_{\text{min}},\,q^2_{\text{max}}]$ as in Refs.~\cite{deBoer:2015boa,deBoer:2018buv,Bause:2019vpr}
\begin{align}
\begin{split}\label{eq:acp_integ}
\tilde{A}_{\text{FB}}&=\frac{1}{\Gamma}\,\left[\int_0^1\,-\,\int_{-1}^0\right]\frac{\text{d}^2\Gamma}{\text{d}q^2\text{d}\cos\theta_\ell}\,\text{d}\cos \theta_\ell = \frac{3\,K_{1c}}{2\Gamma}\,,\\
\tilde{A}_{\text{CP}}&=\frac{\text{d}\Gamma / \text{d} q^2 \,-\,\text{d}\bar\Gamma / \text{d} q^2  }{\Gamma \,+\,\bar\Gamma }=\frac{2\,K_{1ss}+K_{1cc} - 2\,\bar{K}_{1ss}-\bar{K}_{1cc}}{\Gamma + \bar\Gamma}\,.\\
\end{split}
\end{align}
Normalizing to the rate (\ref{eq:acp_integ}) is advantageous as it avoids artificially large asymmetries from a suppressed denominator, such as $d \Gamma/d q^2$, see Fig.~\ref{fig:SMBR}, corresponding to low event rates. It also allows to probe the $q^2$-shape of the numerator.
On the other hand,  cancellations are more efficient when normalizing to $d \Gamma/d q^2$.
Due to the different pros and cons we present results for the CP--asymmetry and the forward-backward asymmetry in both normalizations.
Note, for sufficiently small bins, or sufficiently flat distributions,  the integrals
$\int \tilde A_{\rm CP} dq^2$  and $\int A_{\rm CP} dq^2$, and similarly for the forward-backward asymmetry, of course, coincide.

In the case of the CP--violating forward-backward asymmetry in Eq.~\eqref{eq:afbcp}, we stress that in experiment it might be convenient to measure the sum  and the difference of an angular observable for the decay of particle and antiparticle, and then form the ratio,  Eq.~\eqref{eq:afbcp}.

In the following section \ref{sec:BSM} we investigate in detail the NP sensitivity  of the aforementioned observables.

\section{Sensitivity to Physics Beyond the Standard Model \label{sec:BSM}}

We analyze the NP sensitivity of $c\to u \ell^+ \ell^-$ transitions in $\Lambda_c \to p \ell^+\ell^-$ decays. In Sec.~\ref{sec:constrWCs} we briefly summarize existing constraints on BSM Wilson coefficients. Null tests from the angular distribution are discussed in Sec.~\ref{sec:BSM_nulltests}. We consider lepton universality tests in Sec.~\ref{sec:BSM_lfu}. CP--asymmetries and lepton flavor violating decays are discussed in Sec.~\ref{sec:BSM_cpasym} and \ref{sec:BSM_lfv}, respectively. Sec.~\ref{sec:BSM_dineutrino} provides information on baryon modes with dineutrinos in the final state.

\subsection{Constraints on Wilson coefficients} \label{sec:constrWCs}

Most recent results for lepton flavor conserving and lepton flavor violating Wilson coefficients are presented in~\cite{Gisbert:2020vjx} and based on latest results from the LHCb experiment searching for rare purely leptonic and semileptonic charm meson decays~\cite{Aaij:2020wyk}. Here we use, barring cancellations,
\begin{equation}
\vert C_{7}^{(\prime)}\vert \lesssim 0.3\,,\quad\vert C_{9}^{(\mu)(\prime)}\vert \lesssim 0.9\,,\quad\vert C_{10}^{(\mu)(\prime)}\vert \lesssim 0.8\,,\quad\vert C_{9,\,10}^{(e)(\prime)}\vert \lesssim 4\,,
\end{equation}
for the dimuon and dielectron final state, and
\begin{equation} \label{eq:Klimit}
\vert K_{9,\,10}^{(\mu e)(\prime)}\vert \lesssim 1.6\,,
\end{equation}
for lepton flavor violating (LFV) modes with one electron and one muon in the final state. 
Bounds from high-$p_T$ searches in dilepton tails~\cite{Angelescu:2020uug, Fuentes-Martin:2020lea} are of  the same order of magnitude; 
they are also able to constrain couplings involving $\tau$'s which cannot be kinematically accessed in semileptonic rare charm decays.

\subsection{Angular observables\label{sec:BSM_nulltests}}

\subsubsection{  $A_{\text{FB}}$  and $\tilde{A}_{\text{FB}}$  }
 \begin{figure}[!ht]
   \includegraphics[width=0.49\textwidth]{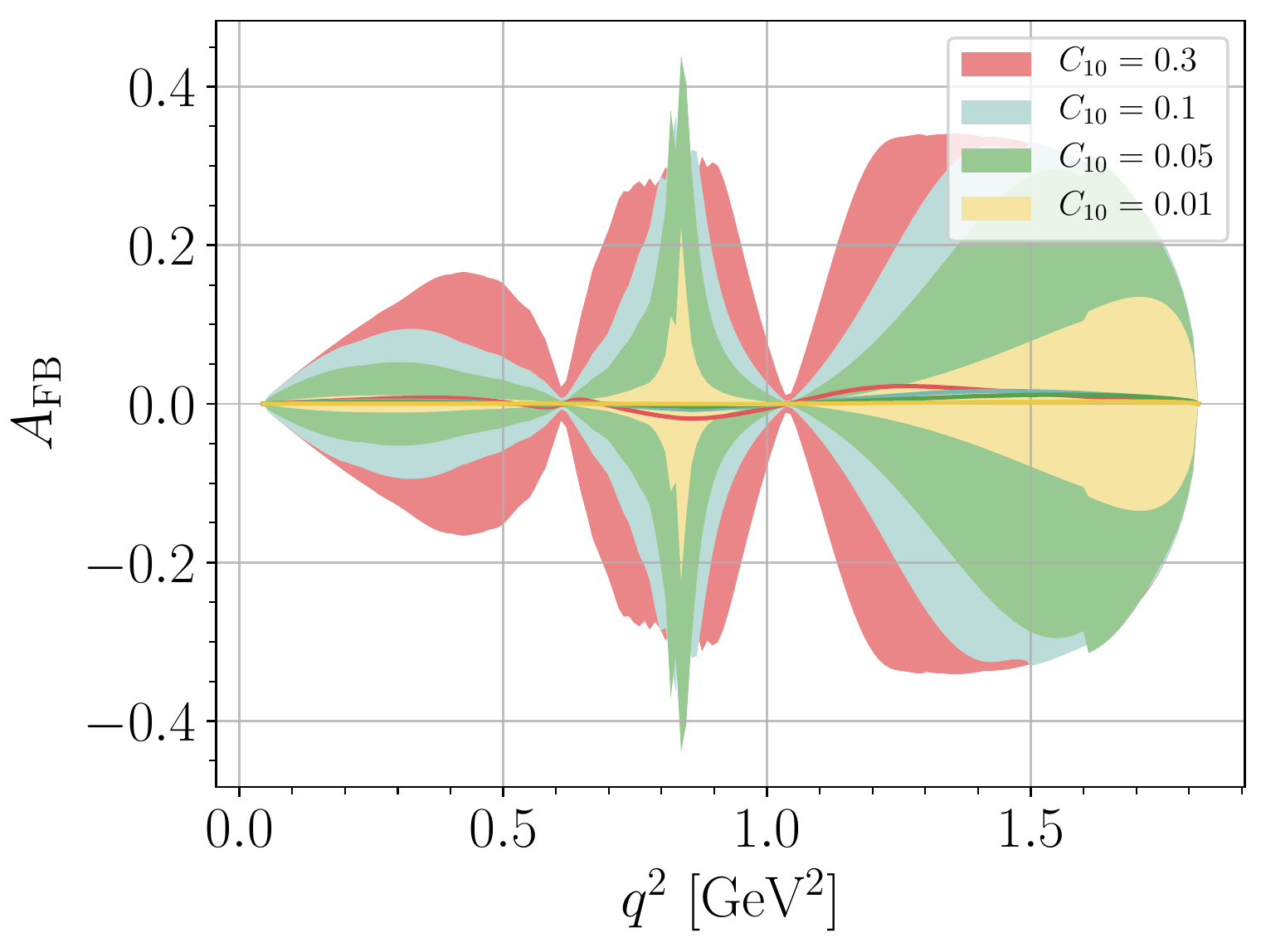}
   \includegraphics[width=0.49\textwidth]{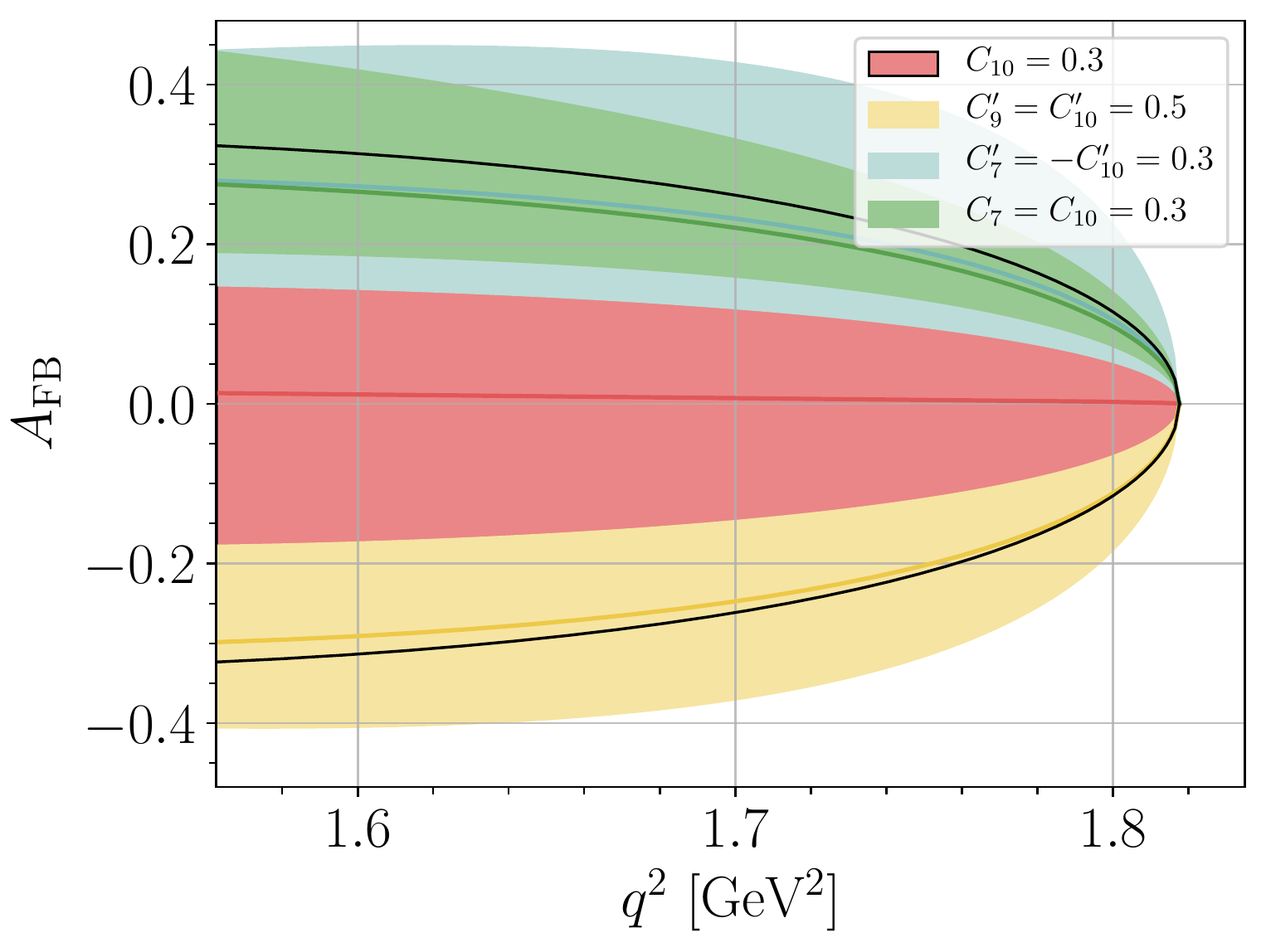}
   \includegraphics[width=0.49\textwidth]{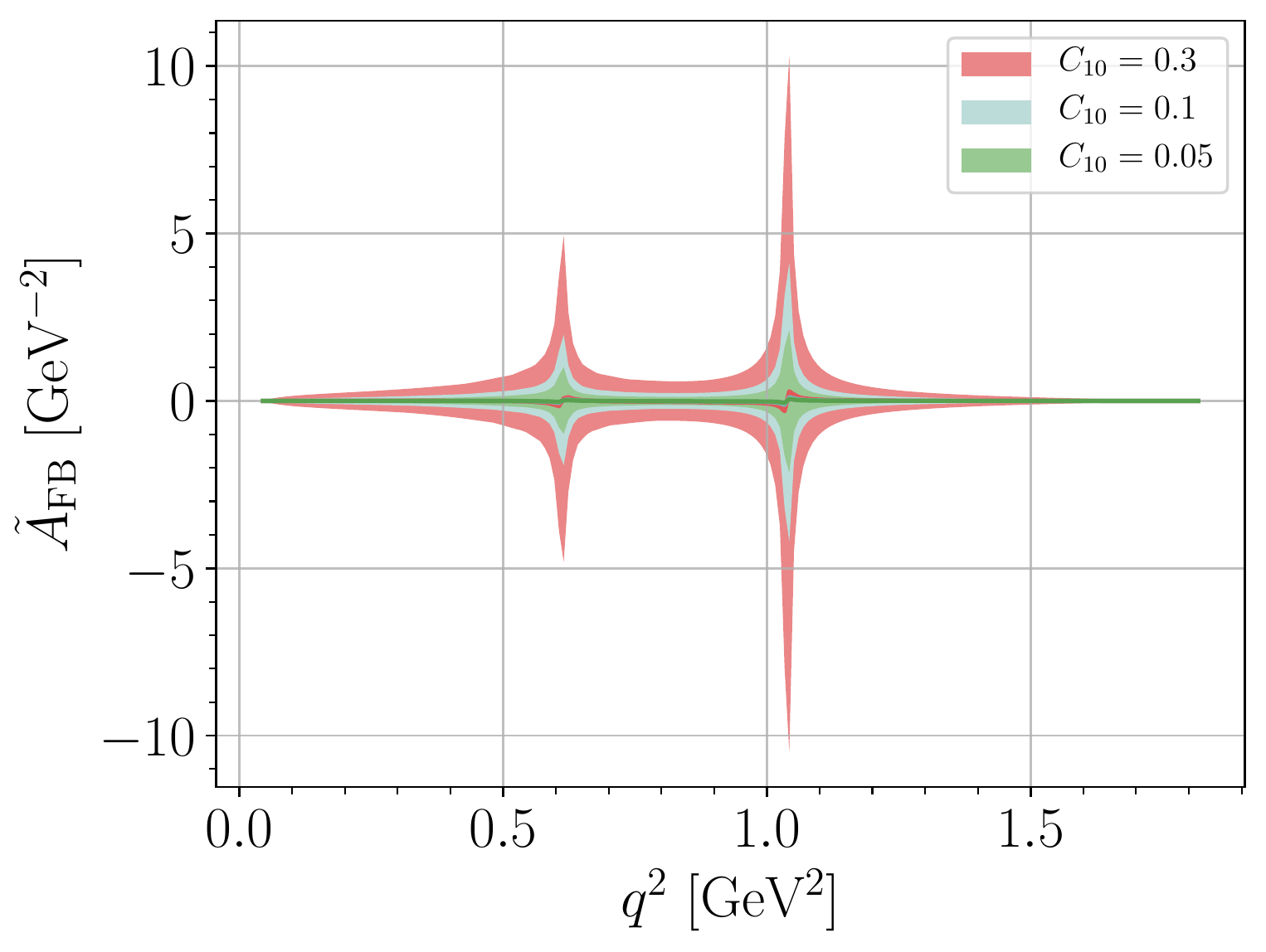}
   \includegraphics[width=0.49\textwidth]{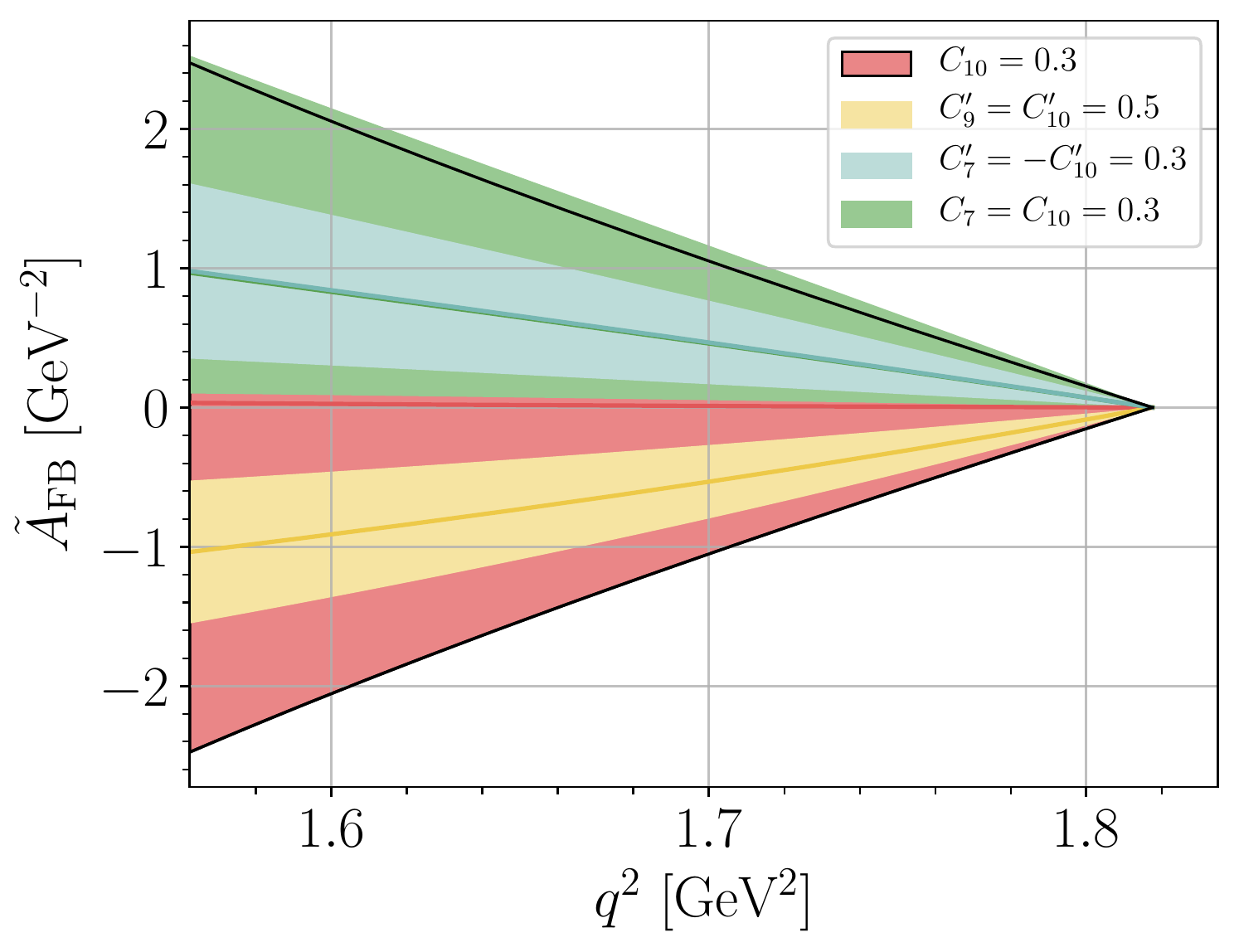}
   \caption{The forward-backward asymmetry $A_{\text{FB}}$ (\ref{eq:afb})  (top) and with normalization to the decay rate  $\tilde{A}_{\text{FB}}$  (\ref{eq:acp_integ}) (bottom) of $\Lambda_c \to p \mu^+  \mu^-$ decays for different values of $C_{10}$ in the full $q^2$-region (left panel) and for various BSM contributions in the  high $q^{2}$ region (right panel), see text.}
          \label{fig:afb}
 \end{figure}

The forward-backward asymmetry in $\Lambda_c \to p \ell^+ \ell^-$ decays 
can exhibit sizable NP effects, see Fig.~\ref{fig:afb}. Here we illustrate the impact of different values of $C_{10}$ on
$A_{\text{FB}}$ (\ref{eq:afb})  (top row) and with normalization to the decay rate,  $\tilde{A}_{\text{FB}}$  (\ref{eq:acp_integ}) (bottom row) 
 in the full $q^2$ region (plots to the left).
 $A_{\text{FB}}$ is markedly suppressed  around  the resonances since here the $q^2$-differential decay rate, which acts as a normalization  (\ref{eq:afb}), is peaking.
Outside the resonances we find that $C_{10}$ even as small as a percent can induce values of $A_{\text{FB}}$ of up to $\mathcal{O}(0.1)$. While this is a clean signal of NP,
hadronic uncertainties are sizable and challenge an extraction of the Wilson coefficients. The inflation of uncertainties is caused by
large cancellations in the differential decay rate.
This situation improves in the high $q^2$ region above the $\phi$-peak (plots to the right), where we show $A_{\text{FB}}$ for $C_{10}$ and in combination with other NP 
coefficients. Only in the former case, illustrated with  $C_{10}=0.3$, $A_{\text{FB}}$ (red region  limited by black solid curve) is symmetric under sign-flip, and 
can vanish for all $q^2$. The other scenarios (green, yellow and blue regions) are shown in the front of the figure and partly cover the red region.

In the lower plots of Fig.~\ref{fig:afb} we show  the forward-backward asymmetry normalized to the integrated decay rate, $\tilde A_{\rm FB}$, defined in (\ref{eq:acp_integ}).
Note the resonance-catalyzed, NP-induced peaks around $m_\rho, m_\omega$ and $m_\phi$. This structure resembles for real-valued NP coefficients the one of  
${\rm Re} \, C_9^R$, shown in Fig.~\ref{fig:perturbative_sm_wcs}.
$\tilde A_{\rm FB}$ has sensitivity to $C_{10}$ as low as a few percent, somewhat weaker than $A_{\text{FB}}$, and in general a more modest uncertainty from varying strong phases. 

Both  $A_{\text{FB}}$  and $\tilde{A}_{\text{FB}}$ vanish  for ${\rm Re} \, C_9^R C_{10}^*=0$, and are invariant under a simultaneous sign-flip of $C_{10}$ and
shift in the resonance phases $\delta_M \to \delta_M +\pi$.
Note that $C_7$ ($C_7^\prime$) can only be probed together with a non-zero  $C_{10}$ ($C_{10}^\prime$).

\subsubsection{$F_L$}

\begin{figure}[!ht]
       \includegraphics[width=0.49\textwidth]{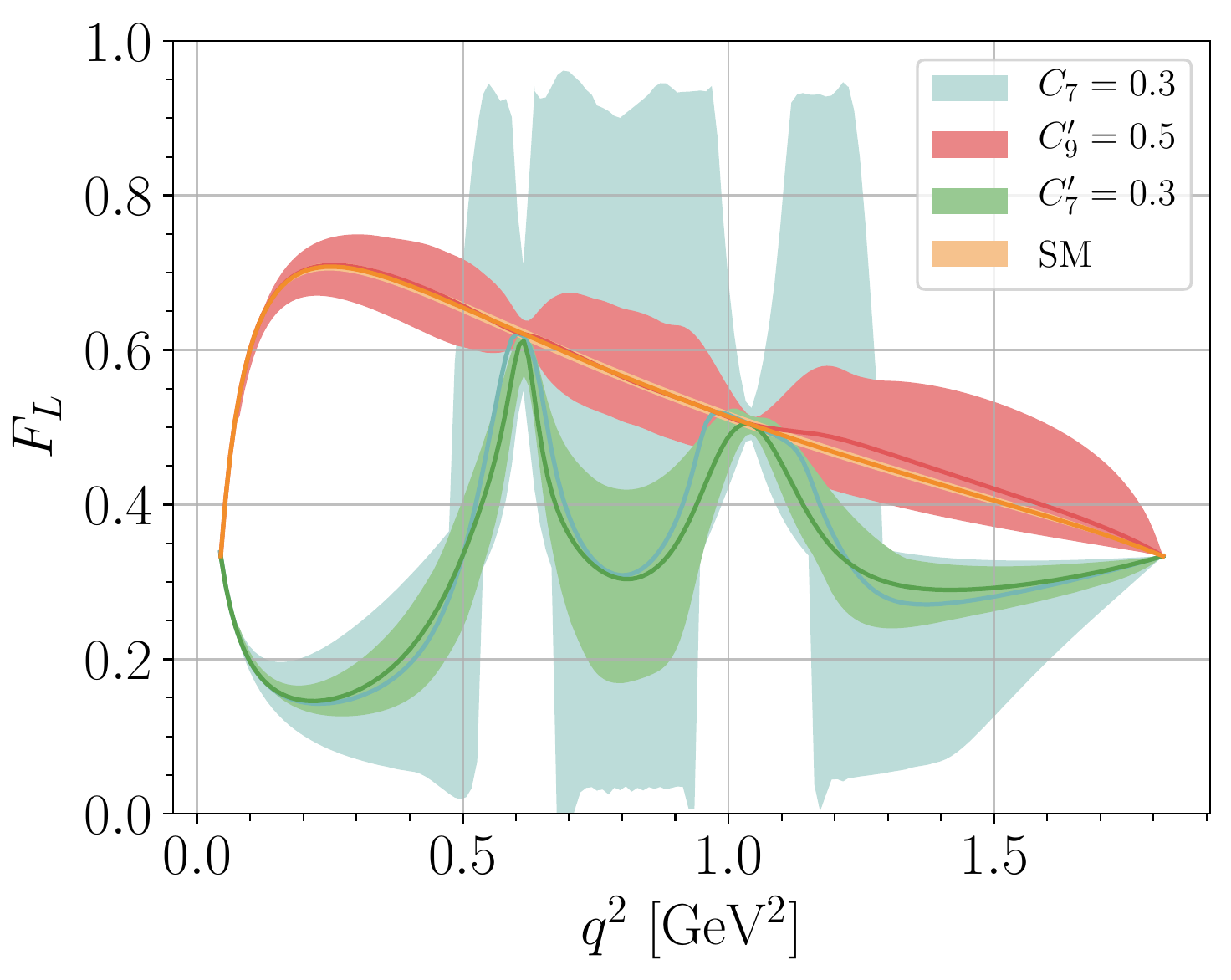}
     \caption{The fraction of longitudinally polarized dimuons $F_L$ of $\Lambda_c \to p \mu^+ \mu^-$ decays for various BSM contributions including theoretical uncertainties from form factors and resonance parameters. The orange band corresponds to the resonant SM contribution.}\label{fig:fl}
 \end{figure}

In contrast to the forward-backward asymmetry the fraction of longitudinally polarized dimuons, $F_L$,  is non-zero  in the SM. The SM prediction of $F_L$, shown in Fig.~\ref{fig:flsm}, is not distinguishable from a NP scenario with only left-handed $C_9,\, C_{10}$ contributions, or  only the right-handed coupling $C_{10}^\prime$ switched on. In  Fig.~\ref{fig:fl} we show the NP sensitivity of $F_L$ to $C_9^\prime$ and  to dipole contributions $C_7^{(\prime)}$. For the former, shown by the red band, the overall $q^2$-shape of $F_L$ follows the resonant-one of the SM (orange line)
modulated with resonance interference contributions and corresponding increased uncertainties from strong phases.
On the other hand, significant deviations from the SM are possible for NP contributions to dipole operators $C_7^{(\prime)} \neq 0$.  Note that the huge uncertainty bands are due to limited knowledge of the strong phases of the resonant contributions, which cancel in the SM case, where predominantly resonance contributions are present, as shown  in Fig.~\ref{fig:flsm}.

To understand the different behavior of NP contributions from $C_{9}^\prime$ and $C_7^{(\prime)}$ in $F_L$, one can use an argument similar to App.~D in Ref.~\cite{Hiller:2013cza}. Using Eqs.~\eqref{eq:2} and~\eqref{eq:36} and considering terms without $m_\ell^2$-suppression, which is suitable for $q^2 \gg 4 m_\ell^2$, $F_L$ can be written as
\begin{equation}\label{eq:fl_leading}
F_L=\frac{L^{11+22}}{U^{11+22}+L^{11+22}}\,.
\end{equation}
Contributions of $C_{9\,,\,10}^\prime$ scale with $m_{\Lambda_c}^2 / q^2$ in $L^{11+22}$, whereas contributions in $U^{11+22}$ have an extra factor two. Thus, the leading $q^2$-dependence from $C_{9\,,\,10}^\prime$-induced contributions to $F_L$ reads
\begin{equation}\label{eq:leading_c9_fl}
F_L\sim \frac{\frac{m_{\Lambda_c}^2}{q^2}}{2+\frac{m_{\Lambda_c}^2}{q^2}}=\frac{1}{\frac{2\,q^2}{m_{\Lambda_c}^2}+1}\,,
\end{equation}
decreasing from $F_L \sim {\cal{O}}(1)$ with increasing $q^2$ towards the zero recoil endpoint where $F_L=1/3$. Contributions with $C_7^{(\prime)}$, on the other hand, receive an additional factor of $2m_c/q^2$ from the photon to lepton pair coupling. The result is an $m_c^2 / q^2$ scaling in $L^{11}$ and an $2 m_c^2\,m_{\Lambda_c}^2 / q^4$ scaling in $U^{11}$, which leads to
\begin{equation}\label{eq:leading_c7_fl}
    F_L\sim \frac{\frac{m_{c}^2}{q^2}}{2\,\frac{m_{c}^2\,m_{\Lambda_c}^2}{q^4}+\frac{m_{c}^2}{q^2}}=\frac{1}{\frac{2\,m_{\Lambda_c}^2}{q^2}+1}\,,
\end{equation}
increasing from $F_L \ll 1$  towards the zero recoil endpoint where $F_L=1/3$.  To understand the behavior for small $q^2$ the $m_\ell^2$-terms matter which dictate
$F_L=1/3$ at maximum recoil.
Together with interference effects and the $q^2$-dependent size of resonant contributions as in Eq.~\eqref{eq:resonances} one obtains the shapes shown  in Fig.~\ref{fig:fl}.

\subsubsection{Towards a global fit}

Unlike in the SM strong phase uncertainties do not cancel in $F_L$ in the presence of NP, and hamper an extraction of Wilson coefficients. 
On the other hand, the
differential branching fraction has high sensitivity to the strong phases. A global fit of future data can therefore benefit from fitting Wilson coefficients
and hadronic phases together. Note that $d {\cal{B}}/dq^2$ and the angular observables $A_{\rm FB}$ and $F_L$ remain invariant under simultaneously flipping of all strong phases in $C_9^R$ and the sign of a NP coefficient. 

\begin{figure}[!ht]
        \includegraphics[width=0.51\textwidth]{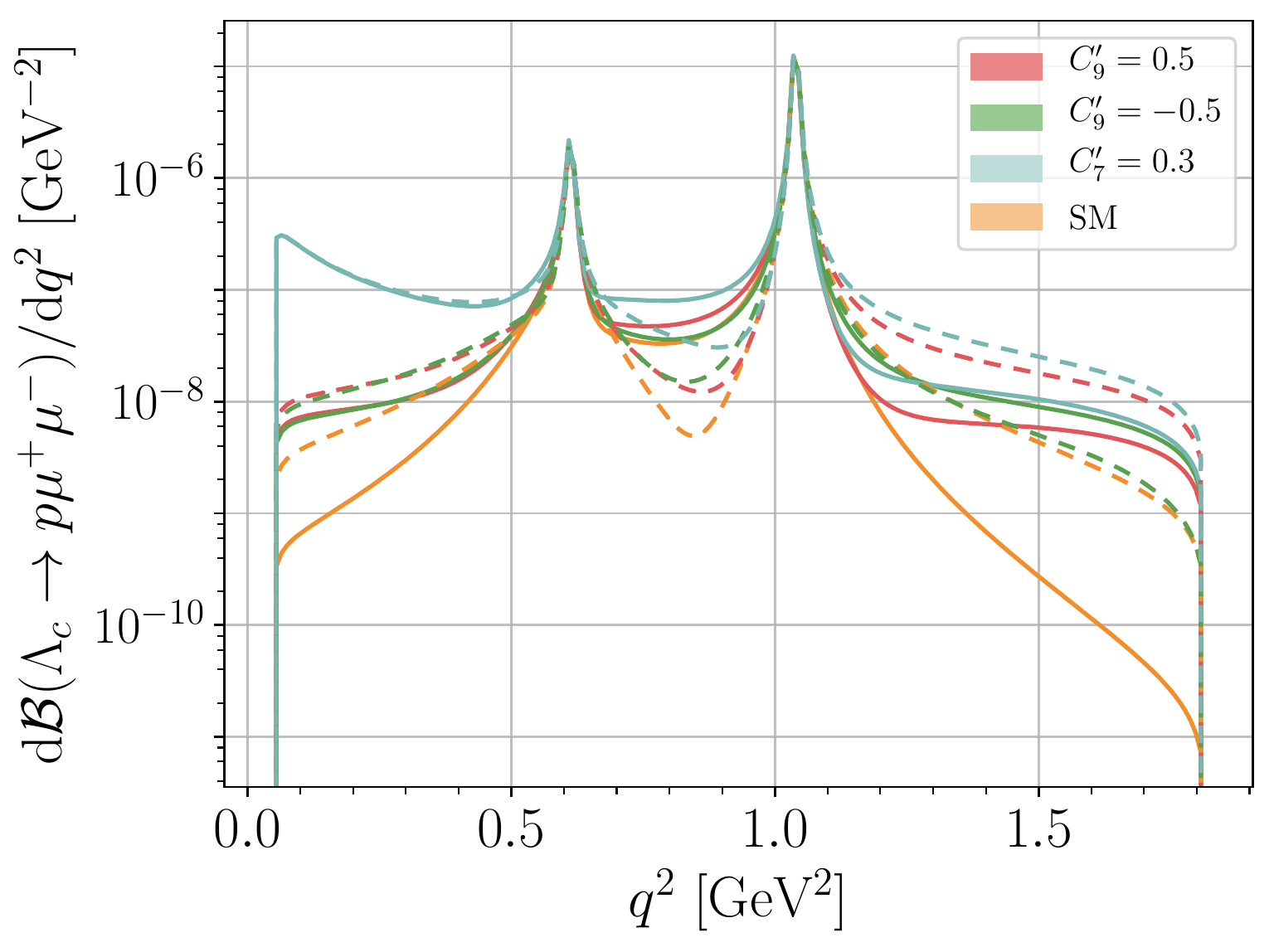}
       \includegraphics[width=0.48\textwidth]{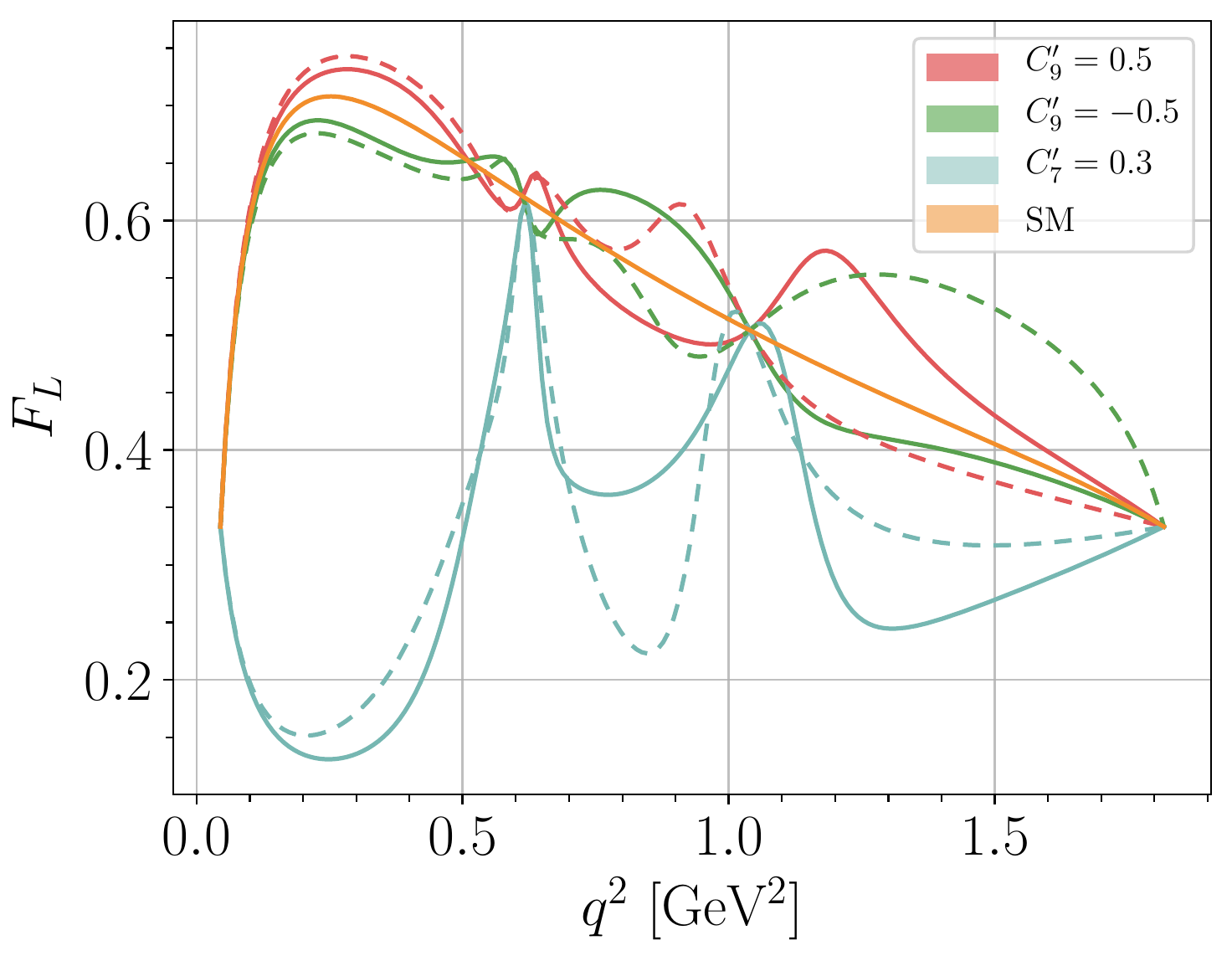}
     \caption{The differential branching fraction (left) and fraction of longitudinally polarized dimuons (right) based on  resonant SM contributions (orange)
     and in NP scenarios with  $C_9^\prime=\pm0.5$ and $C_7^\prime=0.3$. Strong phases are fixed to 
    $\delta_\omega=\delta_\phi=0$ (solid curves)  and $\delta_\omega=0,\,\delta_\phi=\pi$ (dashed curves).  
    Unlike in $d {\cal{B}}/dq^2$, uncertainties due to strong phases cancel in $F_L$ in the SM. See text for details.}
    \label{fig:fl_fixedphases} 
     \end{figure}

 In Fig.~\ref{fig:fl_fixedphases}  we illustrate  the sensitivities and complementarities  between $d {\cal{B}}/dq^2$ (plot to the left)  and  $F_L$ (plot to the right). Shown is the resonant SM in orange and  three NP scenarios.
Strong phases are fixed to 
    $\delta_\omega=\delta_\phi=0$ (solid curves)  and, partly flipped only for clarity, $\delta_\omega=0,\,\delta_\phi=\pi$ (dashed curves).  
Even with this partly flipped strong phases, it can be noticed from $F_L$ that the solid (dashed) curve with $C_9^\prime=0.5$ (red) has similar shape as the
dashed (solid) one for $C_9^\prime=-0.5$ (dashed). 
We recall that $F_L$ is sensitive to right-handed currents, whereas $A_{\rm FB}$ probes $C_{10}$, and therefore both are complementary and important 
 to pin down the type of possible NP.

 The  benchmark $C_7^\prime=0.3$, $\delta_\omega=0,\,\delta_\phi=\pi$  (blue dashed curve) is excluded by the experimental upper limit \eqref{eq:lhcb_upper_limit}. However, since the strong phases remain unknown, $C_7^\prime=0.3$ cannot be  excluded at this point. This highlights the necessity to measure the angular observables and differential branching ratio simultaneously in the same $q^2$ bins, as only then simultaneous fits to Wilson coefficients and resonance parameters can be performed.

\subsection{Lepton universality ratios \label{sec:BSM_lfu}}

The observable $R_p^{\Lambda_c}$ \eqref{eq:R_ratio}  probes electron-muon flavor universality.
We study $R_p^{\Lambda_c}$ in the SM and in  BSM scenarios for the full, low and high $q^2$ regions.
The main source of uncertainty in the branching fraction stems from the resonance contribution, parameterized by  $C_9^R(q^2)$.
We stress that the latter is universal for the muon and the electron mode, and corresponding uncertainties cancel efficiently in the ratio of electron to muon branching fractions.

Predictions for $R_p^{\Lambda_c}$ are given in Tab.~\ref{tab:R_ratios}.
We find near-universality, $1\pm\mathcal{O}(\%)$ for the SM, as lepton mass effects are ${\cal{O}}(m_\ell^2/m_{\Lambda_c}^2)$-suppressed.  A study of higher order QED-corrections is beyond the scope of this work, but may become relevant
once data around this level are available.
We consider  NP scenarios with NP in the muons only to break lepton universality.
Integrating the differential decay rates over the full $q^2$ region we find SM-like values, since the lepton universal resonance contributions dominate. 
On the other hand,
 deviations from the SM can be very large  in the low and high $q^2$ regions, hence, binning is  necessary.
 Hadronic uncertainties are more significant in the high $q^2$ region, where we find values of $R_p^{\Lambda_c}$ of up to $\mathcal{O}(100)$. The theoretical  uncertainties of $R_p^{\Lambda_c}$ originate from  the resonance parameterization and form factors and can be reduced in the future, however, any significant deviation from 1 signals NP.
\begin{table}[!t]
 \centering
  \caption{$R_p^{\Lambda_c}$~\eqref{eq:R_ratio} in the SM and in NP-scenarios with couplings to  muons for different $q^2$--bins as in Eq.~\eqref{eq:q2bins}. Ranges correspond to uncertainties from form factors and resonance parameters. Due to large uncertainties in the high-$q^2$ region, we only provide the order of magnitude for the largest values.}
 \label{tab:R_ratios}
 \begin{tabular}{l||c|c|c|c|c|c|c}
&SM & $\vert C^\mu_9 \vert=0.5$& $\vert C^\mu_{10}\vert=0.5$ & $\vert C^\mu_{9}\vert=\vert C^\mu_{10}\vert=0.5$  & $\vert C^{\prime\mu}_{9}\vert=0.5$ & $\vert C^{\prime\mu}_{10} \vert=0.5$&$\vert C^{\prime\mu}_{9}\vert=\vert C^{\prime\mu}_{10}\vert=0.5$ \\
 \hline
  full $q^2$ & $1.00 \pm \mathcal O(\%)$ &  SM-like & SM-like & SM-like& SM-like& SM-like& SM-like\\
  low $q^2$ & $0.94 \pm \mathcal O(\%)$ & $7.5\ldots20$    &  $4.4\ldots13$     & $11\ldots32$   & $4.6\ldots14$  & $4.4\ldots13$  & $8.2\ldots26$ \\
  high $q^2$ & $1.00 \pm \mathcal O(\%)$ & $\mathcal{O}(100)$    & $\mathcal{O}(100)$     & $\mathcal{O}(100)$ & $\mathcal{O}(100)$  &$\mathcal{O}(100)$ &$\mathcal{O}(100)$ \\
 \end{tabular}
\end{table}

\subsection{CP--asymmetries}\label{sec:BSM_cpasym}

(T--even) CP--asymmetries can be measured in rare charm decays at the resonances~\cite{Fajfer:2012nr}. If there is NP with a CP--violating phase, interference with the resonance contribution leads to a non-vanishing CP--asymmetry. Due to the smallness of CKM-induced CP--violation in $c \to u \ell \ell$ transitions CP--asymmetries are SM null tests.
We study CP--asymmetries in the decay rate, and in the forward-backward asymmetry.

\subsubsection{$A_{\rm CP}$ and $\tilde A_{\rm CP}$}

\begin{figure}[!t]\centering
\includegraphics[width=0.49\textwidth]{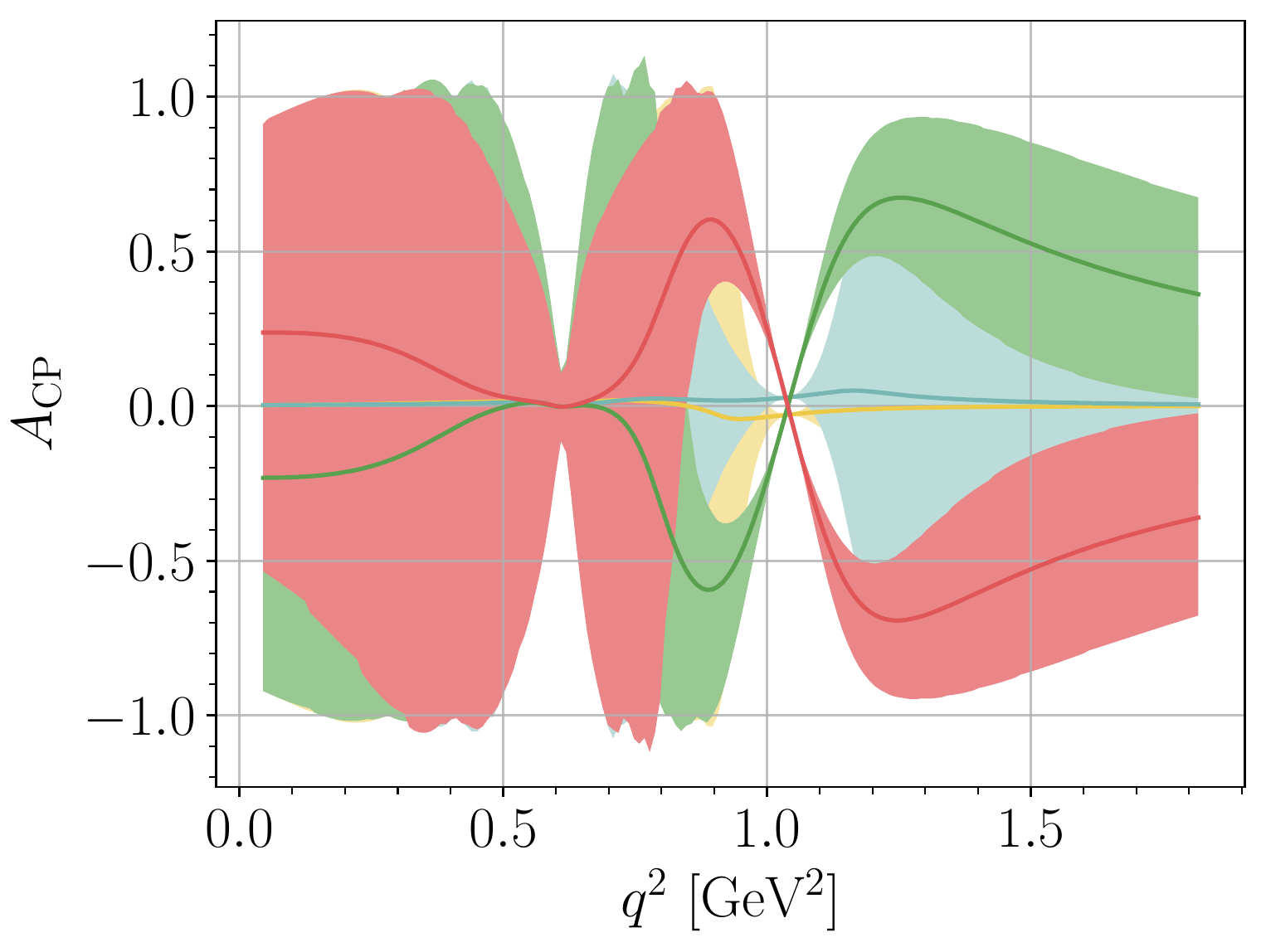}
\includegraphics[width=0.49\textwidth]{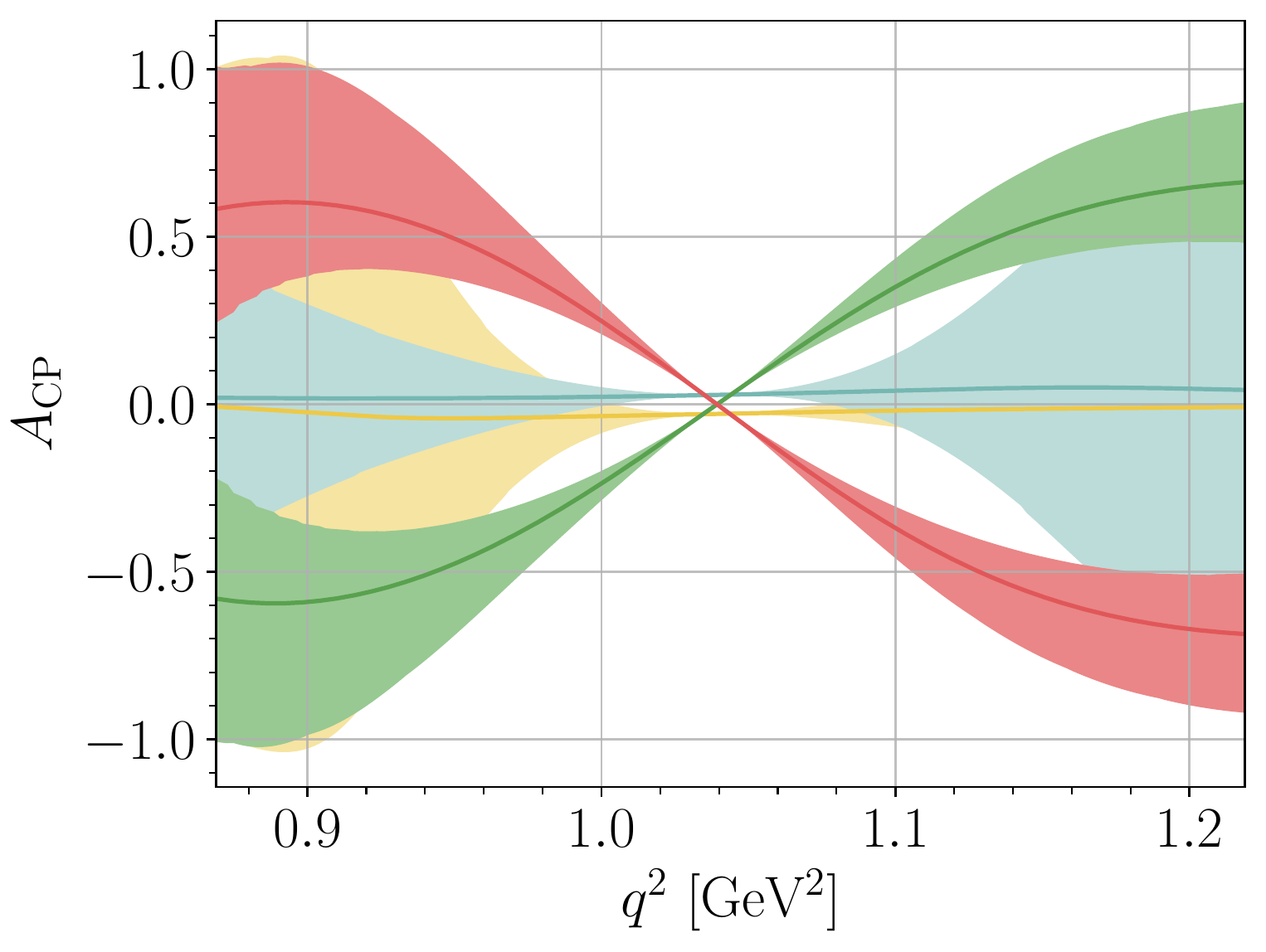}
\includegraphics[width=0.8\textwidth]{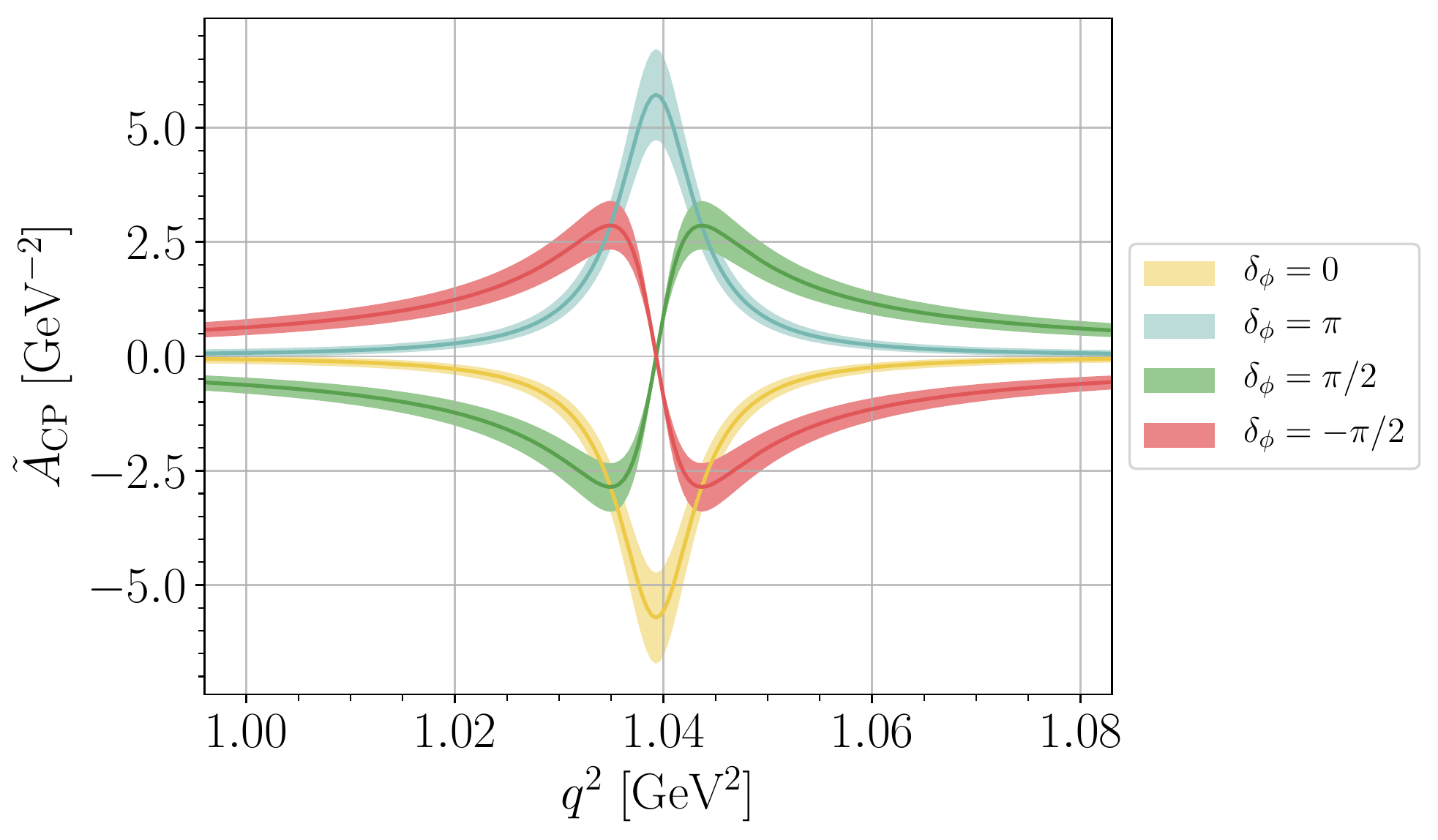}
\caption{The CP--asymmetry of $\Lambda_c \to p \mu^+ \mu^-$ decays for various strong phases in the $C_{9} = 0.5e^{\text{i}\pi / 4}$ scenario with normalization as in Eq.~\eqref{eq:acp} in the full $q^2$ range (left) and the region around the $\phi$ resonance (right). Uncertainties are due to form factors and the strong phase variation in $\delta_\omega$, whereas $\delta_\phi$ is fixed to $0, \pi, \pm\frac{\pi}{2}$ in yellow, blue, green and red, respectively. We also display  $\tilde A_{\rm CP}$  with  alternative normalization \eqref{eq:acp_integ}  for $q^2 \in \left[(m_\phi-5\,\Gamma_\phi)^2,\,(m_\phi+5\,\Gamma_\phi)^2\,\right]$ (bottom).}
\label{fig:acp_phi}
\end{figure}

 In Fig.~\ref{fig:acp_phi} we show the CP--asymmetry \eqref{eq:acp} for  $C_{9} = 0.5e^{i\pi / 4}$ and different strong phases $\delta_\phi=0,\,\pm\frac{\pi}{2},\,\pi$, while varying $\delta_\omega$. 
$A_{\text{CP}}$ vanishes around the resonance peaks, such that smaller values of $A_{\text{CP}}$ coincide with regions with larger $\Lambda_c \to p \ell^+ \ell^-$ event rates.
In Fig.~\ref{fig:acp_phi} (left) the uncertainties are huge around the $\rho,\,\omega$ resonances, because their strong phase is varied in the plot.
On the other hand,  the fixed values of $\delta_\phi$ lead to smaller uncertainties around the $\phi$ resonance.
Improving knowledge on strong phases therefore is important to reduce uncertainties. 
 This can also be seen in the  plot to the right in Fig.~\ref{fig:acp_phi}, which is the same as the  plot to the left  zoomed into the $q^2$-region around the $\phi$.

On the other hand, $\tilde A_{\rm CP}$ with the  integrated decay rate as normalization \eqref{eq:acp_integ} illustrates the behavior of the resonance enhanced CP--asymmetry best
\cite{Fajfer:2012nr, deBoer:2015boa, Bause:2019vpr}, see Fig.~\ref{fig:acp_phi} (lower plot). The $q^2$-integrated denominator
prohibits localized, large cancellations and therefore uncertainty bands are reduced  compared to $A_{\text{CP}}$.
Binning is necessary as,  for instance, CP-asymmetries with  $\delta_\phi=\pm \pi/2$ (green and red) would vanish if integrated symmetrically around $q^2=m_\phi^2$.

\subsubsection{$A^{\rm CP}_{\rm FB}$}

The CP--asymmetry of the forward-backward asymmetry is defined in Eq.~\eqref{eq:afbcp}. Since the forward-backward asymmetry is sensitive to $C_{10}$, one obtains sensitivity to the imaginary part of $C_{10}$, see  \cite{Buchalla:2000sk} for related discussion in $B$-decays. 
In fact, keeping 
only terms with $C_{10}$ and resonance contributions $C_9^R$ in Eq.~\eqref{eq:2}, 
one obtains $K_{1c} \propto P^{12} \propto \text{Re}(C^R_9\,C_{10}^*)$.
In addition, dropping for the moment for simplicity all resonances except the  $\phi$  this implies
\begin{equation}
    \begin{split}
        A_{\text{FB}}^{\text{CP}}=\frac{\text{Im}\,C_9^R}{\text{Re}\,C_9^R}\,\frac{\text{Im}\,C_{10}}{\text{Re}\,C_{10}} \, = \, \frac{(q^2-m_\phi^2)\,\tan \delta_\phi-m_\phi\Gamma_\phi}{q^2-m_\phi^2+m_\phi\Gamma_\phi\,\tan \delta_\phi}\cdot\frac{\text{Im}\,C_{10}}{\text{Re}\,C_{10}}\,.
    \end{split}
    \label{eq:afbcp_resonance}
\end{equation}
Since the strong phase $\delta_\phi$ is unknown, the position of the singularity, induced by the zero in the denominator of Eq.~\eqref{eq:afbcp_resonance}, cannot be predicted. Further singularities arise when all resonance contributions are taken into account. However, the singularities coincide with the zeros in $A_{\text{FB}}$ and joint analysis can help to pin down the strong phases.  In the following it is useful to define  the CP--violating difference and the CP--average individually, respectively, as
\begin{equation}\label{eq:afbcp_individual}
    \begin{split}
       \Delta {A}_{\text{FB}}^{\text{CP}}=\frac{1}{2}\,\left(A_{\text{FB}}+\bar{A}_{\text{FB}}\right)\,,\\
       \Sigma {A}_{\text{FB}}^{\text{CP}}=\frac{1}{2}\,\left(A_{\text{FB}}-\bar{A}_{\text{FB}}\right)\,,
    \end{split}
\end{equation}
where we recall that $A_{\rm FB}$ is CP--odd,  and then Eq.~\eqref{eq:afbcp} is recovered as
\begin{align}\label{eq:afbcp_ratio}
      {A}_{\text{FB}}^{\text{CP}}=\frac{  \Delta {A}_{\text{FB}}^{\text{CP}}}{  \Sigma {A}_{\text{FB}}^{\text{CP}}}\,.
\end{align}
\begin{figure}[!t]\centering
\includegraphics[width=0.49\textwidth]{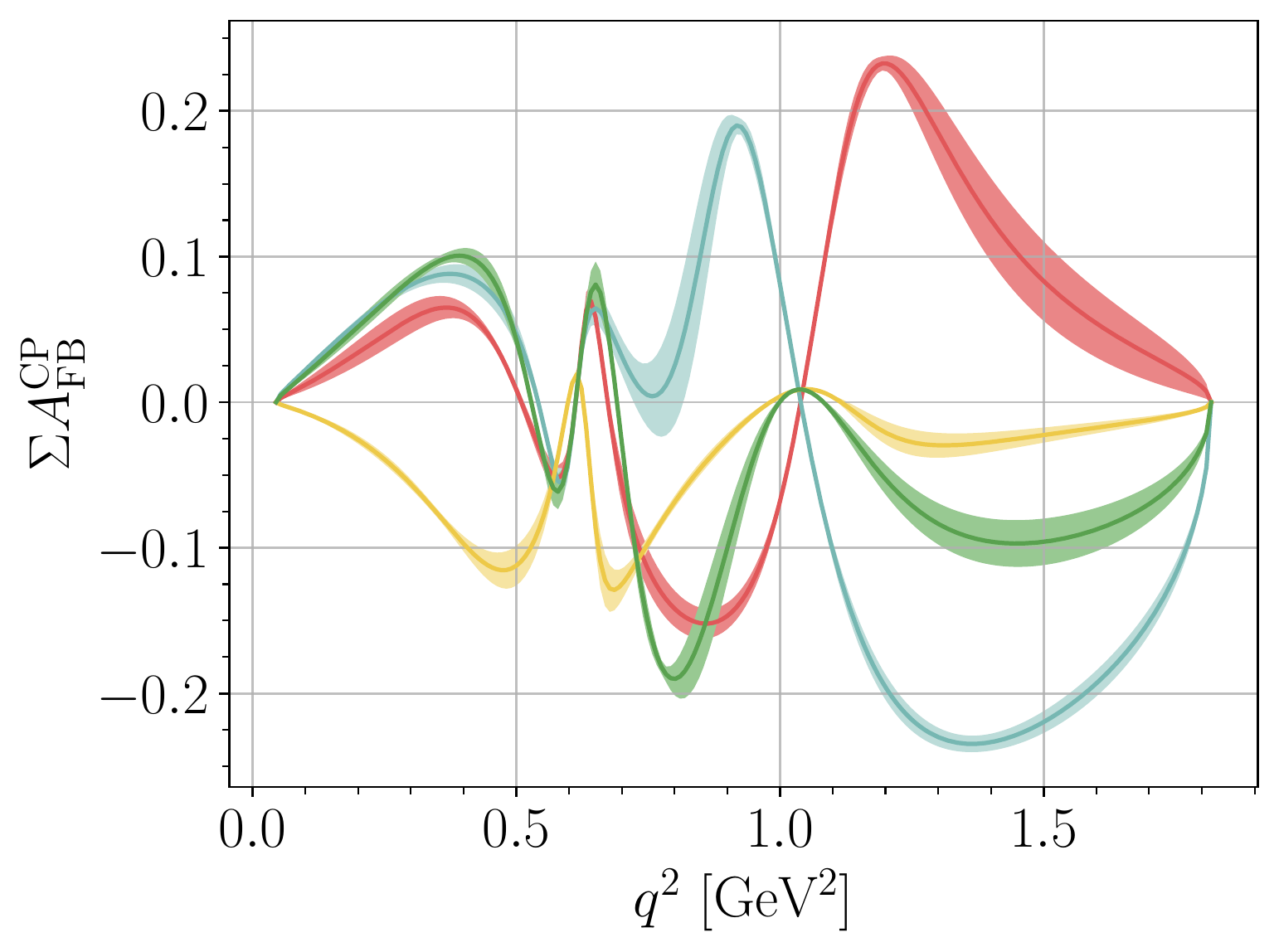}
\includegraphics[width=0.49\textwidth]{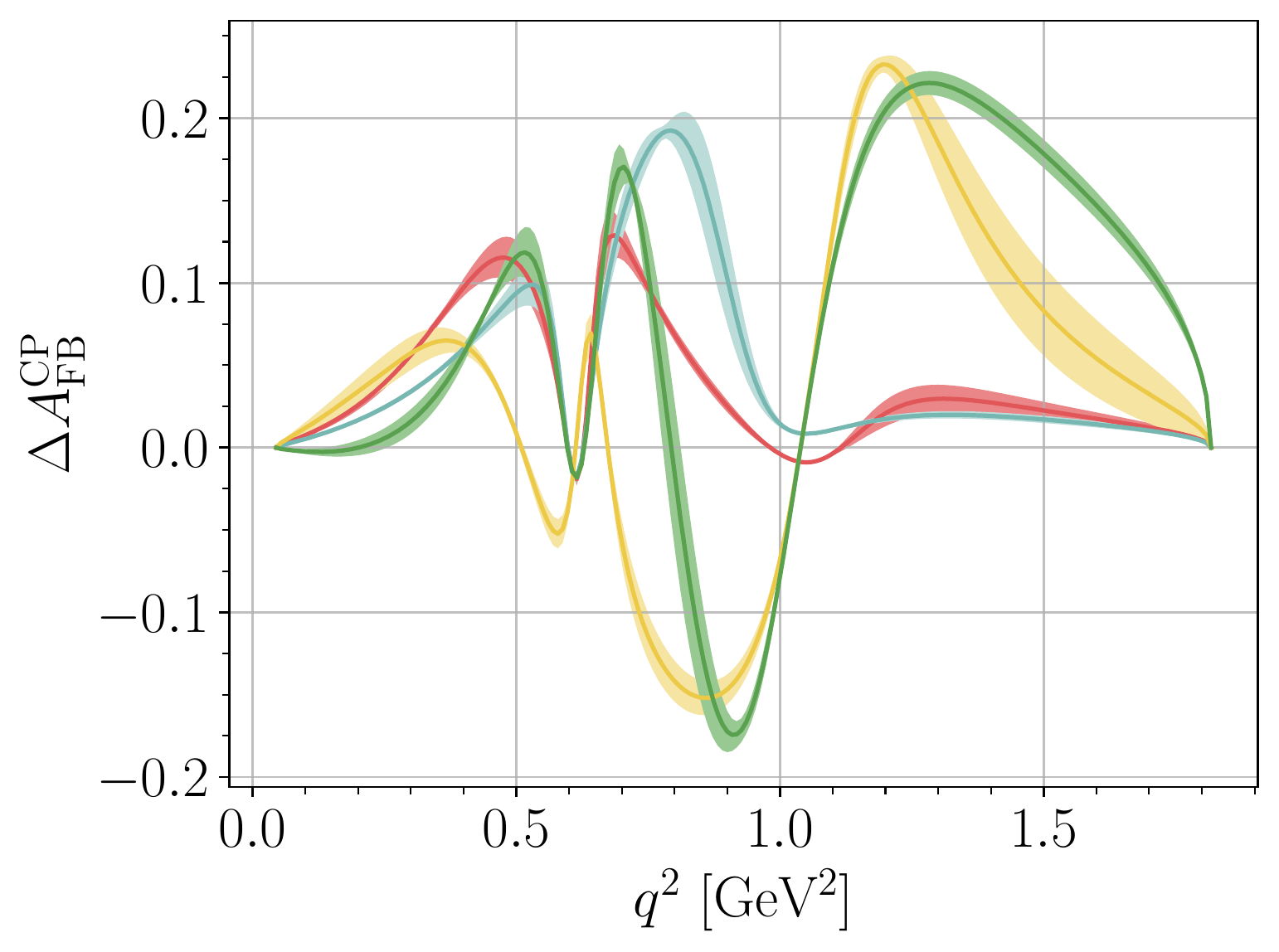}
\includegraphics[width=0.8\textwidth]{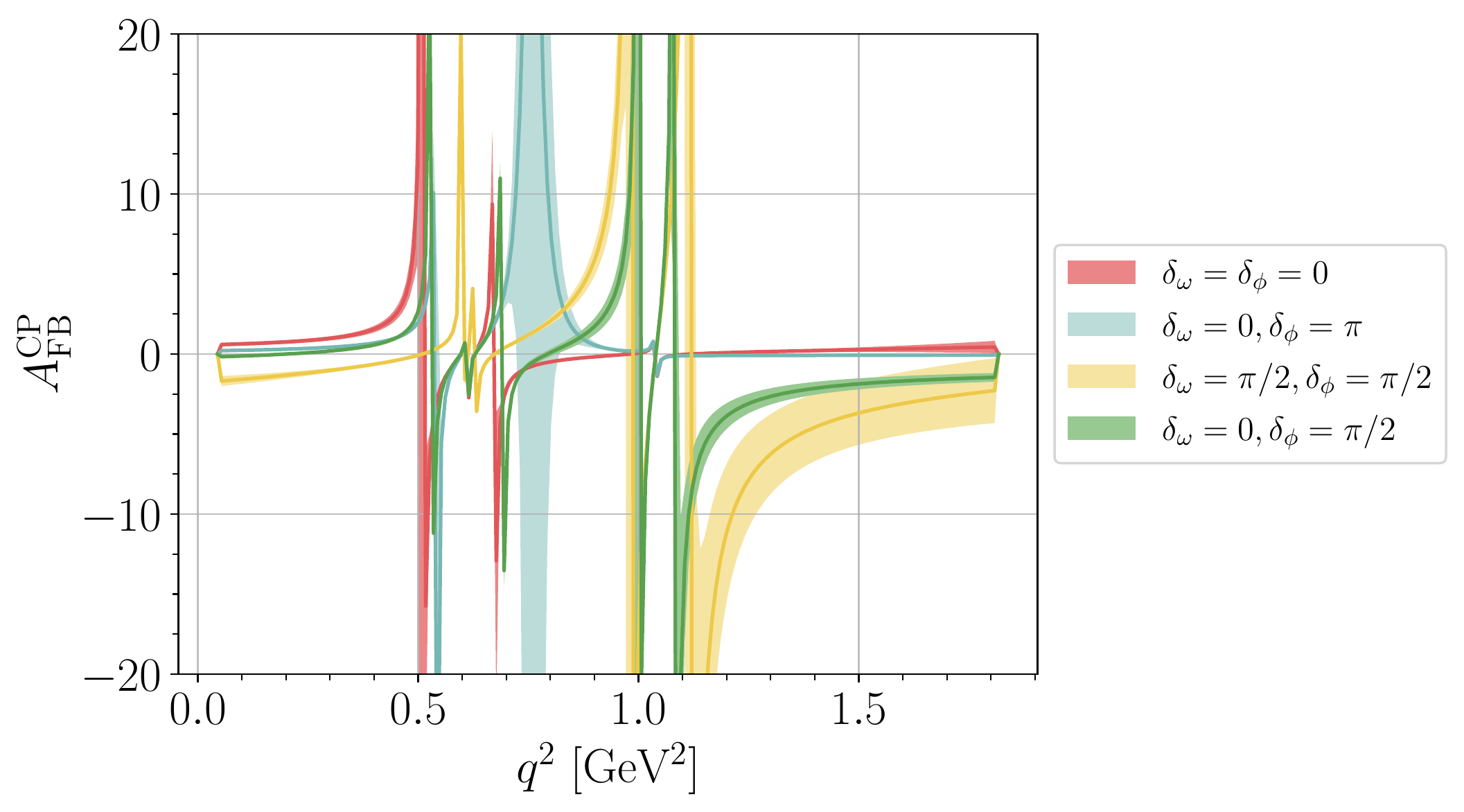}
\caption{CP--asymmetry of  the forward-backward asymmetry in $\Lambda_c \to p \mu^+ \mu^-$ decays for $C_{10} = 0.5\,e^{\text{i}\pi/4}$ scenario and  different, fixed strong phases. The upper plots  show the CP--average (left) and CP--difference (right) corresponding to Eq.~\eqref{eq:afbcp_individual}, the lower plot corresponds to their ratio Eqs.~\eqref{eq:afbcp},~\eqref{eq:afbcp_ratio}.}
\label{fig:acpfb_phi}
\end{figure}
In  Fig.~\ref{fig:acpfb_phi} we show $ \Sigma {A}_{\text{FB}}^{\text{CP}}$ (upper left) and $ \Delta {A}_{\text{FB}}^{\text{CP}}$ (upper right), as well as their ratio (lower plot) corresponding to Eqs.~\eqref{eq:afbcp},~\eqref{eq:afbcp_ratio}  for  $C_{10}=0.5\,e^{i\pi/4}$ and  different  combinations of strong phases. As already anticipated, the  singularities in ${A}_{\text{FB}}^{\text{CP}}$  (lower plot)  coincide  with the zeros in $ \Sigma {A}_{\text{FB}}^{\text{CP}}$ (upper left plot).
Similar to the results in Sec.~\ref{sec:BSM_nulltests} we learn that binned measurements are crucial to disentangle not only the NP contributions but also the size of the strong phases. Although the singularities  make it impossible to give predictions for integrated values of $A_{\text{FB}}^{\text{CP}}$, binned measurements are highly sensitive to the size of strong phases.
Note that there is no sensitivity in  ${A}_{\text{FB}}^{\text{CP}}$ to $C_{10}^\prime$, as its contribution does not involve requisite strong phases.

\subsection{Lepton flavor violation}\label{sec:BSM_lfv}

For the discussion of LFV decays with $c\to u\ell^-\ell^{\prime+}$ ($\ell\neq\ell^\prime$) the following effective Hamiltonian is defined,
\begin{equation} \label{eq:LFV}
\mathcal H_\text{eff}^{\rm LFV} \supset -{4G_F \over \sqrt2} {\alpha_e \over 4\pi} \sum_{i=9,10} \left( K_i^{(\ell\ell^\prime)} O_i^{(\ell\ell^\prime)} + K_i^{\prime\,(\ell\ell^\prime)} O_i^{\prime\,(\ell\ell^\prime)} \right) \,,
\end{equation}
where the $K_i^{(\prime)}$ denote Wilson coefficients and the operators  read as
\begin{align}
O_9^{(\ell\ell^\prime)} & =  (\ubar_L \gamma_\mu c_L) (\lbar \gamma^\mu \ell^\prime) \,, \quad\quad
O_9^{\prime\,(\ell\ell^\prime)} = (\ubar_R \gamma_\mu c_R) (\lbar \gamma^\mu \ell^\prime) \,, \nonumber \\
O_{10}^{(\ell\ell^\prime)} & =  (\ubar_L \gamma_\mu c_L) (\lbar \gamma^\mu \gamma_5 \ell^\prime) \,, \quad\quad
O_{10}^{\prime\,(\ell\ell^\prime)} = (\ubar_R \gamma_\mu c_R) (\lbar \gamma^\mu  \gamma_5 \ell^\prime) \,.
\label{eq:lfv_operators}
\end{align}
Note that there is no $O_7^{(\prime)}$ contribution since the photon does not couple to different lepton flavors.
For $\ell \neq \ell'$ the angular observables in Eq.~\eqref{eq:angl_distr} contain additional terms due to non-vanishing lepton amplitudes~\cite{Das:2019omf}.
The modified angular observables in terms of the lepton masses $m_{\ell}$ and $m_{\ell'}$ are given in Appendix~\ref{app:lfv}.
We present distributions and branching ratios  summing both lepton charge combinations, defining, for instance,
$\mathcal{B}(\Lambda_c\to p \mu^\pm e^\mp)=\mathcal{B}(\Lambda_c\to p \mu^-e^+)+\mathcal{B}(\Lambda_c\to p \mu^+ e^-)$.

\begin{figure}[t!]\centering
\includegraphics[width=0.6\textwidth]{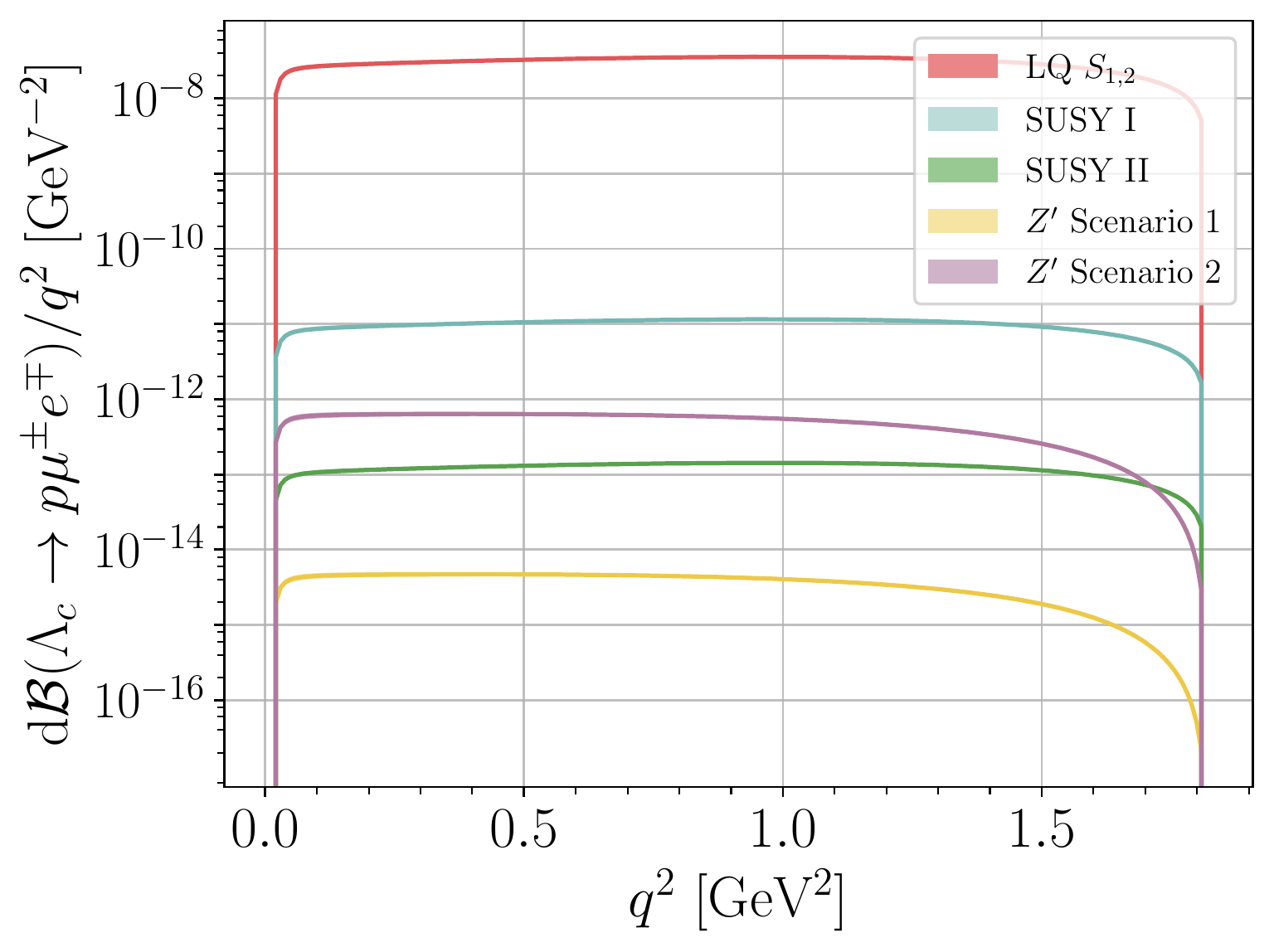}
\caption{The differential branching fraction for $\Lambda_{c} \to p\mu^{\pm}e^{\mp}$ with lepton flavor violating NP contributions in a LQ scenario ($K'_9 = K'_{10} = 0.5$), SUSY scenarios with and without R-parity violation ($K_9 = - K_{10} = 0.009$ and $K_9 = - K_{10} = 0.001$) and two different $Z^\prime$ scenarios ($K_9 = K'_9 = - K_{10} = -K'_{10} = 1.4\cdot 10^{-4}$ and $K_9 = K'_9 = - K_{10} = -K'_{10} = 2.3\cdot 10^{-4}$). The BSM benchmarks are adapted from Ref.~\cite{Bause:2019vpr}. The error bands include form factor uncertainties. }
\label{fig:lfv_br}
\end{figure}

Fig.~\ref{fig:lfv_br} shows the differential branching fraction for the decay~$\Lambda_{c} \to p\mu^{\pm}e^{\mp}$ in different lepton flavor violating BSM scenarios.
Branching ratios up to $\mathcal{O}(10^{-8})$ are possible in LQ scenarios, whereas SUSY models only reach few$\times\mathcal{O}(10^{-12})$ and $Z^\prime$ models are at most at $\mathcal{O}(10^{-13})$. LFV branching ratios are clean SM null tests, as no resonance pollution and no SM background exists. Bounds on LFV Wilson coefficients are 
also of special interest for combinations with flavor summed dineutrino branching ratios, see Sec.~\ref{sec:BSM_dineutrino}.
The presented NP benchmarks are sufficiently below the $90\,\%$ CL upper limits from BaBar \cite{Lees:2011hb} which read
\begin{align}
\begin{split}
\mathcal{B}_{\text{Babar}}(\Lambda_c\to p \mu^-e^+)<9.9\cdot 10^{-6}\,,\\
\mathcal{B}_{\text{Babar}}(\Lambda_c\to p \mu^+e^-)<19\cdot 10^{-6}\,.
\end{split}
\end{align}

Model-independent upper limits on LFV branching ratios based on Eq.~\eqref{eq:Klimit} read
\begin{align}
\label{eq:UL-LFV}
\begin{split}
\mathcal{B}(\Lambda_c\to p  \mu^\pm e^\mp)\lesssim 8.2 \cdot 10^{-7}\,,\\
\mathcal{B}(\Xi_c^{+} \to \Sigma^{+} \mu^\pm e^\mp )\lesssim1.6 \cdot 10^{-6}\,, \\
\mathcal{B}(\Xi_c^{0} \to \Sigma^{0} \mu^\pm e^\mp )\lesssim2.6 \cdot 10^{-7}\,, \\
\mathcal{B}(\Xi_c^{0} \to \Lambda^{0} \mu^\pm e^\mp )\lesssim1.2 \cdot 10^{-7}\,, \\
\mathcal{B}(\Omega_c ^0\to \Xi^0 \mu^\pm e^\mp )\lesssim1.4 \cdot 10^{-6}\,.
\end{split}
\end{align}

\subsection{Dineutrino modes}\label{sec:BSM_dineutrino}

Null test opportunities  for rare charm decays with dineutrinos modes were recently 
presented~\cite{Bause:2020xzj, Bause:2020auq}. The experimental signature includes missing energy and requires a clean environment such as an $e^+e^-$-machine. 
Due to the strong GIM suppression and absence of  QCD resonances, $c \to u \nu \bar \nu$ branching ratios in the SM are well beyond any experimental reach in the foreseeable future. 
 Interestingly,  NP can induce sizable rates which are in reach of present flavor facilities Belle II \cite{Kou:2018nap}, BES III \cite{Ablikim:2019hff}, and beyond \cite{Abada:2019lih}.
Moreover,  the  achievable upper limits on the branching ratios depend on charged lepton flavor, therefore, a search for rare charm dineutrino modes not only tests the SM,
it tests at the same time lepton universality (LU), or  charged lepton flavor conservation (cLFC)~\cite{Bause:2020xzj, Bause:2020auq}.
The latter originates from the SU(2)$_L$-link between left-handed charged leptons and neutrinos.
Since the neutrino flavors are not tagged,
the upper limits on dineutrino branching ratios are obtained as an incoherent sum of all final state flavors
\begin{align}
\mathcal{B}\left( c \to u \,\nu \bar \nu\right)=\sum_{i,j} \mathcal{B}\left( c\to u \,\nu_i  \bar \nu_j\right)  \, ,
\end{align}
which are constrained by charged dilepton data\cite{Bause:2020xzj, Bause:2020auq}. Note, the PMNS-matrix drops out from this link, and that the upper limits are data-driven, and can be improved. Bounds on dilepton couplings, including $\tau$'s,  exist from high-$p_T$-data~\cite{Angelescu:2020uug, Fuentes-Martin:2020lea}.
The strongest upper limit arises in models with LU; since the muon bounds are the strongest they dictate the limit.
Relaxing LU but within cLFC branching ratio limits increase. In the most general situation, denoted below by "general", that is, LFV, the maximum limits
allowing only for light left-handed neutrinos are obtained.

Predictions for rare charm baryon modes read \cite{Bause:2020xzj}
\begin{equation} \label{eq:la-nunu}
\begin{split}
\mathcal{B}(\Lambda_c^+\to p\nu\bar\nu) \lesssim 1.8\cdot10^{-6}\quad &\text{(LU)}\,,\\
\mathcal{B}(\Lambda_c^+\to p\nu\bar\nu) \lesssim 1.1\cdot10^{-5}\quad &\text{(cLFC)}\,,\\
\mathcal{B}(\Lambda_c^+\to p\nu\bar\nu) \lesssim 3.9\cdot10^{-5}\quad &\text{(general)}\,,
\end{split}
\end{equation}
and
\begin{equation} \label{eq:xi-nunu}
\begin{split}
\mathcal{B}(\Xi_c^+\to \Sigma^+\nu\bar\nu) \lesssim 3.6\cdot10^{-6}\quad &\text{(LU)}\,,\\
\mathcal{B}(\Xi_c^+\to \Sigma^+\nu\bar\nu) \lesssim 2.1\cdot10^{-5}\quad &\text{(cLFC)}\,,\\
\mathcal{B}(\Xi_c^+\to \Sigma^+\nu\bar\nu) \lesssim 7.6\cdot10^{-5}\quad &\text{(general)}\,.
\end{split}
\end{equation}
Following  Ref.~\cite{Bause:2020xzj} we further obtain
\begin{equation} \label{eq:xi-nunu}
\begin{split}
\mathcal{B}(\Xi_c^0\to \Sigma^0\nu\bar\nu) \lesssim 6.2\cdot10^{-7}\quad &\text{(LU)}\,,\\
\mathcal{B}(\Xi_c^0\to \Sigma^0\nu\bar\nu) \lesssim 3.6\cdot10^{-6}\quad &\text{(cLFC)}\,,\\
\mathcal{B}(\Xi_c^0\to \Sigma^0\nu\bar\nu) \lesssim 1.3\cdot10^{-5}\quad &\text{(general)}\,,
\end{split}
\end{equation}
\begin{equation} \label{eq:xi-nunu}
\begin{split}
\mathcal{B}(\Xi_c^0\to \Lambda^0\nu\bar\nu) \lesssim 2.7\cdot10^{-7}\quad &\text{(LU)}\,,\\
\mathcal{B}(\Xi_c^0\to \Lambda^0\nu\bar\nu) \lesssim 1.5\cdot10^{-6}\quad &\text{(cLFC)}\,,\\
\mathcal{B}(\Xi_c^0\to \Lambda^0\nu\bar\nu) \lesssim 5.6\cdot10^{-6}\quad &\text{(general)}\,,
\end{split}
\end{equation}
and
\begin{equation} \label{eq:omeg-nunu}
\begin{split}
\mathcal{B}(\Omega_c^0\to \Xi^0\nu\bar\nu) \lesssim 3.4\cdot10^{-6}\quad &\text{(LU)}\,,\\
\mathcal{B}(\Omega_c^0\to \Xi^0\nu\bar\nu) \lesssim 1.9\cdot10^{-5}\quad &\text{(cLFC)}\,,\\
\mathcal{B}(\Omega_c^0\to \Xi^0\nu\bar\nu) \lesssim 7.1\cdot10^{-5}\quad &\text{(general)}\,.
\end{split}
\end{equation}
A measurement above  a bound based on a lepton flavor symmetry implies the violation of the latter in the charged lepton sector. For instance,  $\mathcal{B}(\Lambda_c\to p\nu\bar\nu) > 1.8\cdot10^{-6}$ implies lepton universality violation, $R_p^{\Lambda_c}\neq1$, and $\mathcal{B}(\Lambda_c\to p\nu\bar\nu) > 1.1\cdot10^{-5}$ implies non-vanishing signals in LFV modes.

\section{Conclusions}\label{sec:summary}

We present a comprehensive analysis of rare charm baryon decays mediated by  $c\to u \ell^+ \ell^-$, 
$c\to u \mu^\pm e^\mp$ and  $c\to u \nu \bar \nu$
 transitions. The three-body decays $\Lambda_c\to p \ell \ell^{(\prime)}$ and friends, $\Xi_c^{+} \to \Sigma^{+}  \ell \ell^{(\prime)}$, $\Xi_c^{0} \to \Sigma^{0}  \ell \ell^{(\prime)}$,
  $\Xi_c^{0} \to \Lambda^{0}  \ell \ell^{(\prime)}$ and  $\Omega_c^0 \to \Xi^0 \ell \ell^{(\prime)}$,  are sensitive to a rich set of new physics phenomena, such as breakings of symmetries that hold approximately in the SM.
 In particular,
 
 \begin{itemize}
 \item $A_{\rm FB}$, shown in   Fig.~\ref{fig:afb}, probes $C_{10}$, leptonic axial-vector couplings, which are protected in the SM by the GIM-mechanism.
  $A_{\rm FB}$ allows to probe $C_{10}$ at order $10^{-2}$, about two orders below the present bound (\ref{eq:Klimit}).
 If normalized to the decay rate, shown in the lower left plot in Fig.~\ref{fig:afb}, NP signals are enhanced on resonance.
 
 \item $F_L$ probes electromagnetic dipole contributions $C_{7}^{(\prime)}$, see  Fig.~\ref{fig:fl}. 
 In addition, $F_L$ is  sensitive to NP effects from right-handed currents with leptonic vector coupling $C_{9}^{\prime}$.
 $F_L$ is therefore complementary to $A_{\rm FB}$.
 The reach in $F_L$ is order percent and few percent in $C_{7}^{(\prime)}$ and $C_{9}^{\prime}$, respectively, allowing to significantly improve existing limits on the 
 Wilson coefficients (\ref{eq:Klimit}).

  \item CP--violation: The CP--asymmetry in the decay rates  shown in Fig.~\ref{fig:acp_phi} is sensitive to CP-violation in left-handed $C_9$ and right-handed currents 
  $C_9^\prime$, and dipole couplings $C_7^{(\prime)}$. The forward-backward CP--asymmetry
 $A_{\rm FB}^{\rm CP}$, shown in Fig.~\ref{fig:acpfb_phi}, probes ${\rm Im} \, C_{10}$.

 \item
  Lepton flavor universality can be probed in lepton flavor specific modes, using ratios of branching fractions,  $R^{\Lambda_c}_p$,
  comparing dimuon to dielectron final states (\ref{eq:R_ratio}).  NP effects in $R^{\Lambda_c}_p$  can be very large in the high $q^2$ region, see Tab.~\ref{tab:R_ratios}.
   Lepton flavor universality can also be probed with branching ratios into dineutrino final states \cite{Bause:2020xzj, Bause:2020auq}, which are null tests of the SM.
  Data-driven, upper limits assuming universality (LU) are given in  (\ref{eq:la-nunu}) - (\ref{eq:omeg-nunu}).

 \item Lepton flavor violation can be searched for  in  $e^\pm \mu^\mp$ final states, which allow for sizable, model-independent  rates (\ref{eq:UL-LFV}), 
 also in concrete SM extensions, such as leptoquarks, shown in Fig.~\ref{fig:lfv_br}.
Charged lepton flavor violation can also be probed with branching ratios into dineutrino final states \cite{Bause:2020xzj, Bause:2020auq},
Upper limits assuming flavor conservation (cLFC)  are provided in  (\ref{eq:la-nunu}) - (\ref{eq:omeg-nunu}).

 \end{itemize}
 
 Even more Wilson coefficients can be probed if more than one receives NP contributions. Such a broad program  calls for a fit that should be performed once data are available, where we stress the importance of $q^2$-binning and, more general, probing the shapes of distributions.
 We further suggest to simultaneously fit process-dependent hadronic phases along with the Wilson coefficients.  The differential branching ratios within
 the $\omega,\rho$ and the $\phi$ peaks are particularly sensitive to such phases, see  Fig.~\ref{fig:fl_fixedphases}  and also  \cite{deBoer:2018buv}.
 Such a global $c \to u$  fit should also include rare $D$-meson measurements. Recall that angular distributions in $D \to P \ell^+ \ell^-$, $P=\pi,K$, modes are SM null tests which probe  semileptonic (pseudo-)scalar and tensor operators \cite{Bause:2019vpr}, different from those in (\ref{eq:operators}). Rare charm baryon decays also involve both types of coupling combinations $C_i + C_i^\prime$ and $C_i - C_i^\prime$. The importance of rare charm baryon decays is amplified as the decays into vector mesons, such as $D \to \rho \ell^+ \ell^-$ have, for  kinematic reasons, 
 essentially no high-$q^2$-region above the $\phi$.
 Rare charm baryon modes therefore complement  searches with $D$-meson decays, and together, they advance our understanding of the up-type flavor sector.

\section*{Acknowledgments}
We thank Nico Adolph, Dominik Mitzel and Roman Zwicky for useful discussions. This work is supported by the \textit{Studienstiftung des Deutschen Volkes} (MG) and  in part by the \textit{Bundesministerium f\"ur Bildung und Forschung} (BMBF) under project number
 05H21PECL2 (GH).

\appendix
\section{Standard Model Wilson Coefficients}\label{app:WCs}
For completeness  we give the SM Wilson coefficients using Refs.~\cite{deBoer:thesis, deBoer:2017way, deBoer:2015boa}. The calculation is done in a different basis than the one utilized in Eq.~\eqref{eq:Heff}. Instead, 
\begin{align}
\mathcal{H}^{\text{weak}}_{\text{eff}}\Big\vert_{m_W\geq\mu > m_b}&=-\frac{4\,G_F}{\sqrt{2}}\,\sum_{q\in\{d,\,s,\,b\}}\,V_{cq}^*V_{uq}\,\left(\tilde{C}_1(\mu)P_1^{(q)}+\tilde{C}_2(\mu)P_2^{(q)}\right)\,,\\
\mathcal{H}^{\text{weak}}_{\text{eff}}\Big\vert_{m_W > \mu\geq m_c}&=-\frac{4\,G_F}{\sqrt{2}}\,\sum_{q\in\{d,\,s\}}\,V_{cq}^*V_{uq}\,\left(\tilde{C}_1(\mu)P_1^{(q)}+\tilde{C}_2(\mu)P_2^{(q)}+\sum_{i=3}^{10}\tilde{C}_i(\mu)P_i\right)\,,
\end{align}
\begin{align}
 &P_1^{(q)}=(\bar u_L\gamma_{\mu_1}T^aq_L)(\overline q_L\gamma^{\mu_1}T^ac_L)\,,\\
 &P_2^{(q)}=(\bar u_L\gamma_{\mu_1}q_L)(\overline q_L\gamma^{\mu_1}c_L)\,,\\
 &P_3=(\bar u_L\gamma_{\mu_1}c_L)\sum_{\{q:m_q<\mu\}}(\overline q\gamma^{\mu_1}q)\,,\\
 &P_4=(\bar u_L\gamma_{\mu_1}T^ac_L)\sum_{\{q:m_q<\mu\}}(\overline q\gamma^{\mu_1}T^aq)\,,\\
 &P_5=(\bar u_L\gamma_{\mu_1}\gamma_{\mu_2}\gamma_{\mu_3}c_L)\sum_{\{q:m_q<\mu\}}(\overline q\gamma^{\mu_1}\gamma^{\mu_2}\gamma^{\mu_3}q)\,,\\
 &P_6=(\bar u_L\gamma_{\mu_1}\gamma_{\mu_2}\gamma_{\mu_3}T^ac_L)\sum_{\{q:m_q<\mu\}}(\overline q\gamma^{\mu_1}\gamma^{\mu_2}\gamma^{\mu_3}T^aq)\,,\\
 &P_7=\frac e{g_s^2}m_c\left(\bar u_L\sigma^{\mu_1\mu_2}c_R\right)F_{\mu_1\mu_2}\,,\\
 &P_8=\frac 1{g_s}m_c\left(\bar u_L\sigma^{\mu_1\mu_2}T^ac_R\right)G^a_{\mu_1\mu_2}\,,\\
 &P_9=\frac{e^2}{g_s^2}(\bar u_L\gamma_{\mu_1}c_L)\left(\overline \ell\gamma^{\mu_1}\ell\right)\,,\\
 &P_{10}=\frac{e^2}{g_s^2}(\bar u_L\gamma_{\mu_1}c_L)\left(\overline \ell\gamma^{\mu_1}\gamma_5 \ell\right)\,.
\end{align}

Contributions to Wilson coefficients  are calculated systematically at (partly) NNLL order in QCD, \textit{e.g.} expanded in powers of $\alpha_s(\mu)$ as
\begin{align}
\tilde{C}_i(\mu)=\tilde{C}_i^{(0)}(\mu)+\frac{\alpha_s(\mu)}{4\pi}\,\tilde{C}_i^{(1)}(\mu)+\left(\frac{\alpha_s(\mu)}{4\pi}\right)^2\,\tilde{C}_i^{(2)}(\mu)+\mathcal{O}(\alpha_s^3(\mu))\,,
\end{align}
where the most important noteworthy properties of the calculation are listed below.
\begin{itemize}
  \item $\tilde{C}_1$ and $\tilde{C}_2$ are calculated at the electroweak scale and evaluated down to the $b$ quark mass within renormalization improved perturbation theory.
  \item At the $b$-quark mass scale, a five to four flavor matching is performed, as the $b$ quark is integrated out. Due to the matching,  operators other than $P_{1,2}$ are induced.
  \item After matching at $\mu_b$, all contributions are evolved  to the charm mass scale, where $\mu_c\in \left[\frac{m_c}{\sqrt{2}},\, \sqrt{2}\,m_c\right]$ is taken as an scale uncertainty.  In addition the $W$ and $b$ mass scales are also varied by a factor 2 to estimate further scale uncertainties.
  \item $\tilde{C}_{10}$ is zero in the SM to all orders.
\end{itemize}
The calculation is performed in \textit{Python} with  numerical results summarized in Tab.~\ref{tab:WCs}, where the first three rows show the complete NNLO result for the three different charm mass scales and the fourth and fifth row the resulting central value and uncertainty, respectively. Our values are in perfect agreement\footnote{There is a  sign error in Table 2.2 of Ref.~\cite{deBoer:thesis} in $\tilde{C}_6^{(1)}$. }
with Refs.~\cite{deBoer:thesis, deBoer:2017way, deBoer:2015boa}.

\begin{table}[!ht]
 \centering
  \caption{SM Wilson coefficients $\tilde{C}_j$ for $j\in [1,\, 10]$ up to NNLO accuracy. In the first three rows we show values for the charm mass scale $\mu=m_c$ and for scale variations of $\mu=\sqrt{2}\,m_c$ and $\mu=m_c/\sqrt{2}$, respectively. The last two rows explicitly give the central values and uncertainties, which are used  in this work.}
 \label{tab:WCs}
 \begin{tabular}{l||c|c|c|c|c|c|c|c|c|c}
& $j=1$ & $j=2$ & $j=3$ & $j=4$ & $j=5$ & $j=6$ & $j=7$ & $j=8$ & $j=9$ & $j=10$ \\
\hline
$\tilde{C}_j(\mu=m_c)$& -0.6402&1.0349&-0.0084&-0.0953& 0.0005& 0.0009& 0.0038&-0.0024&-0.0133& 0.0000\\
$\tilde{C}_j(\mu=\sqrt{2}\,m_c)$& -0.5159 & 1.0257&-0.0037&-0.0583& 0.0002& 0.0002& 0.0020&-0.0013&-0.0083&  0.0000\\
$\tilde{C}_j(\mu=m_c/\sqrt{2})$& -0.7902& 1.0431&-0.0184&-0.1649& 0.0011& 0.0031& 0.0071&-0.0046&-0.0214 & 0.0000\\
\hline
\hline
$\tilde{C}_j$&-0.64& 1.035& -0.008 & 0.095 & 0.0005& 0.0009 & 0.004 & -0.002& -0.013 & 0.0000 \\
$\Delta{\tilde{C}}_j$ & 0.14 & 0.009 & 0.007 & 0.05& 0.0005 & 0.0015 & 0.003& 0.002& 0.006 &0.0000\\
 \end{tabular}
\end{table}

\begin{figure}[!t]\centering
\includegraphics[width=0.49\textwidth]{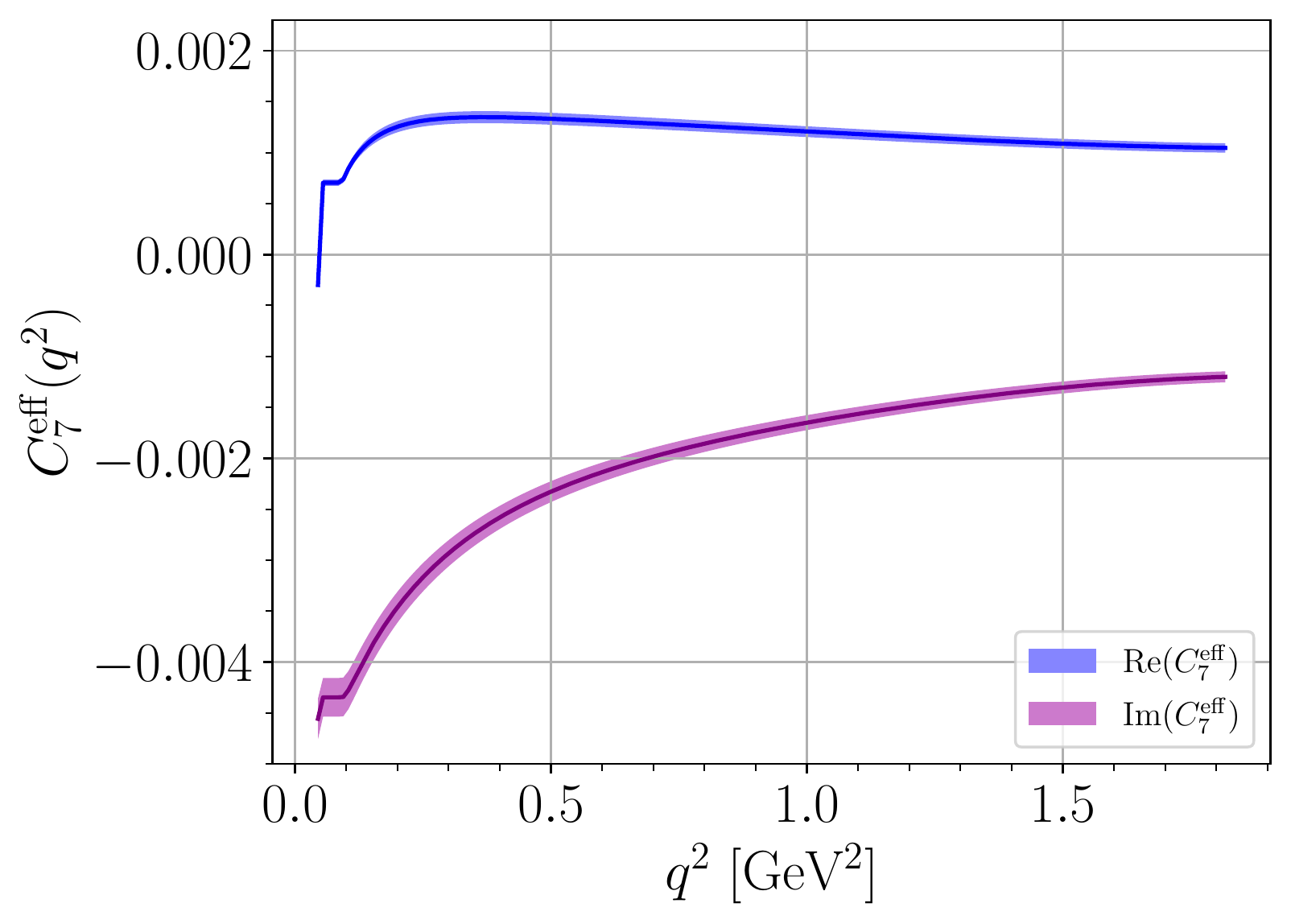}
\includegraphics[width=0.49\textwidth]{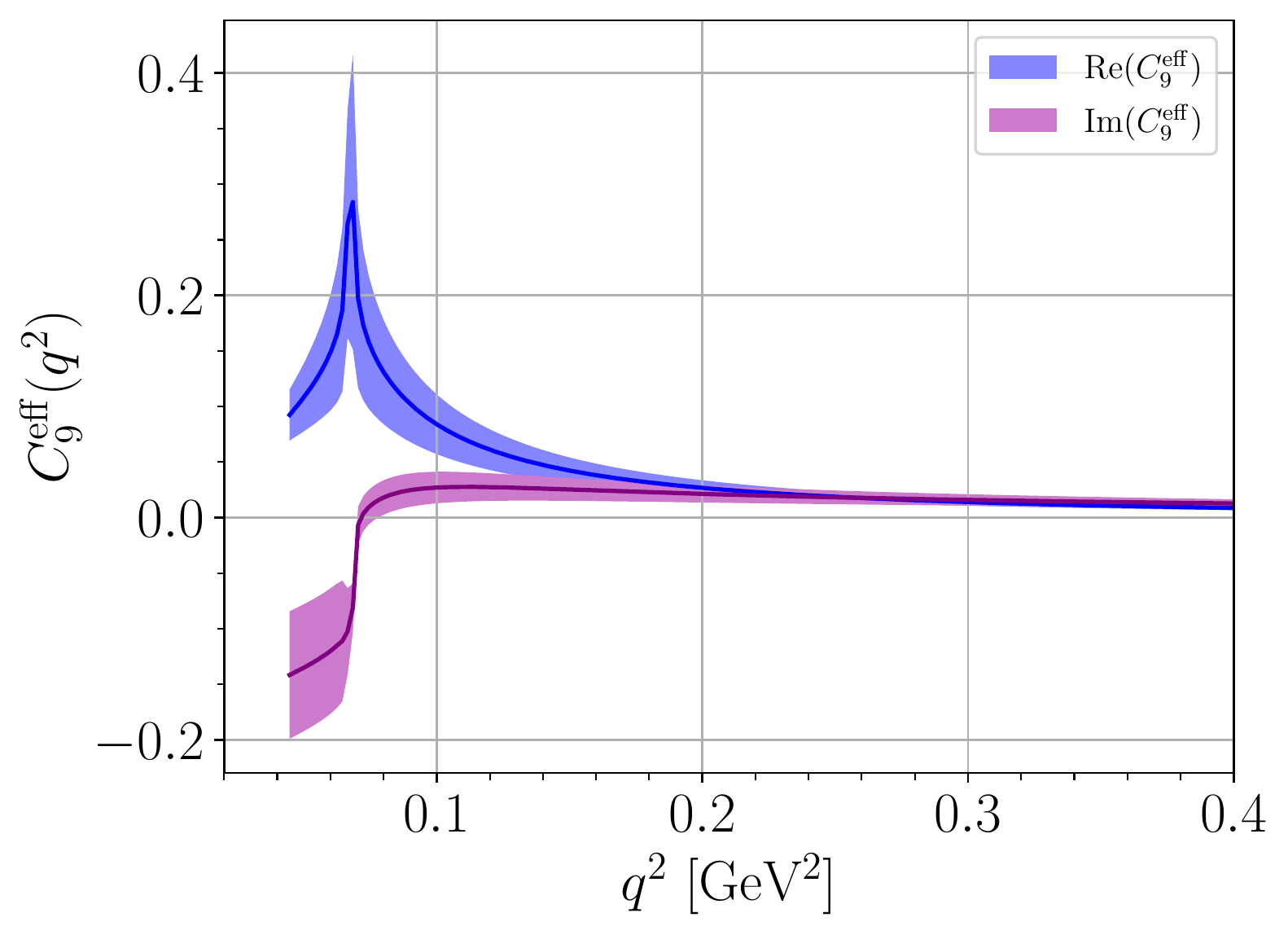}
\includegraphics[width=0.49\textwidth]{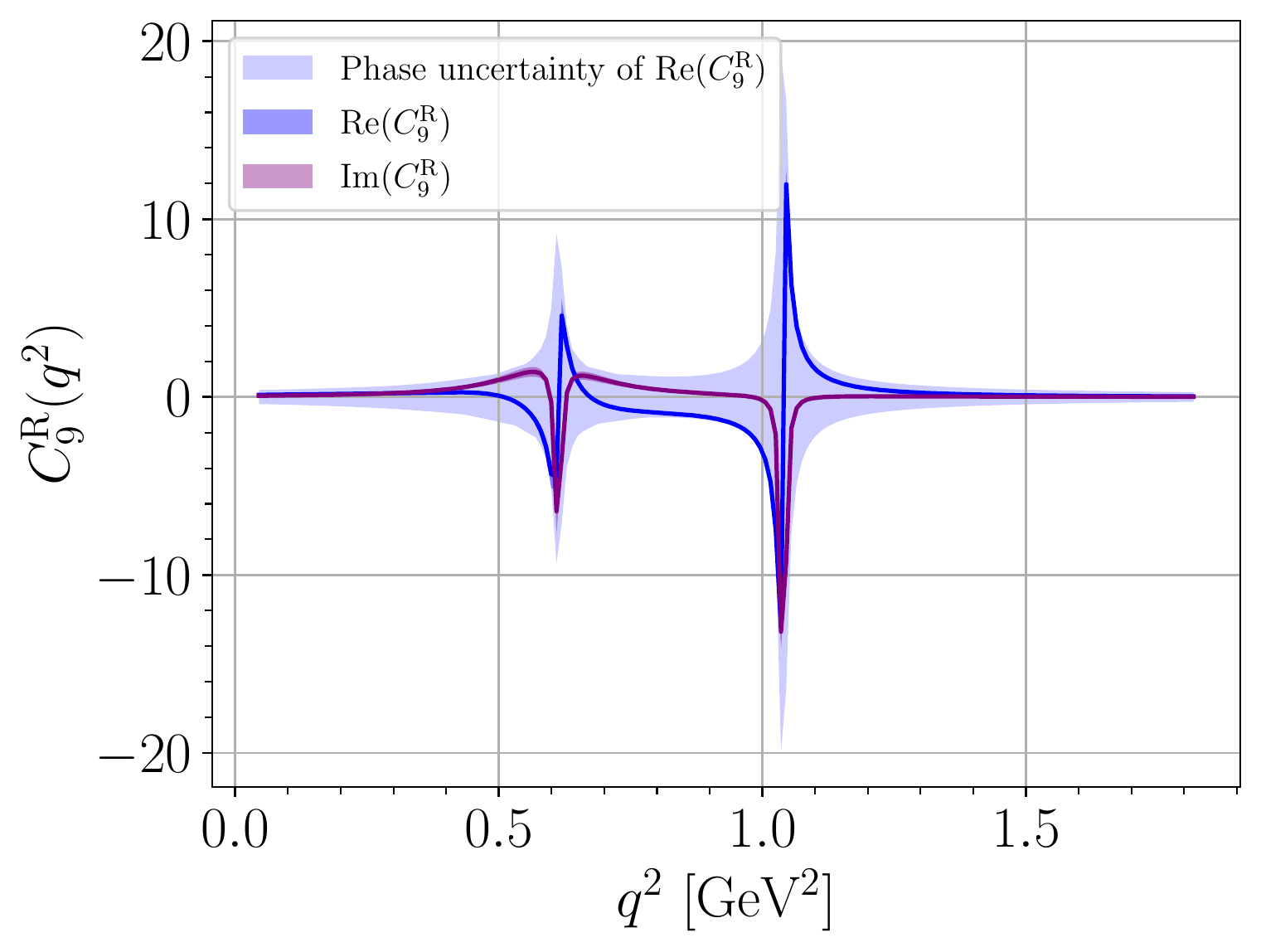}
\caption{Real (blue) and imaginary (lilac) parts of perturbative contributions to $C^{\text{eff}}_7$ (upper left), $C_9^{\text{eff}}$ (upper right) and the resonant $C^\mathrm{R}_9$ (bottom) in the SM as functions of $q^2$. $C_9^{\text{eff}}(q^2)$ remains essentially constant for $q^2 \gtrsim 0.4 \, \mbox{GeV}^2$ and is therefore not shown. The main source of uncertainty in the perturbative contributions is due to the scale uncertainty $\mu_c$ and indicated as bands. In the resonant case the solid lines show  $C^\mathrm{R}_9$ for fixed  phases $\delta_\phi = \delta_\omega = 0$, whereas the shaded (blue) area 
corresponds to ${\rm Re} C^\mathrm{R}_9$ including uncertainties from varying  phases.
The corresponding uncertainty band for ${\rm Im} C^\mathrm{R}_9$ is not shown because it is very similar.}
\label{fig:perturbative_sm_wcs}
\end{figure}

The Wilson coefficients $\tilde{C}_j$ are absorbed into ``effective`` coefficients $C^{\text{eff}(q)}_7$ and $C^{\text{eff}(q)}_9$, $q=d, \, s$, up to 
NNLO, see Ref.~\cite{deBoer:2015boa, deBoer:thesis, deBoer:2017way} for details.
In terms of the phenomenological basis in Eq.~\eqref{eq:Heff},
\begin{align}
C^{\text{eff}}_{7,\,9}(q^2)=\sum_{q=d,\,s}\,V_{cq}^*V_{uq}\frac{4\pi}{\alpha_s(\mu_c)}\,C^{\text{eff}(q)}_{7,\,9}(\mu_c,\, q^2),
\end{align}
which define the perturbative SM contributions. They are shown with their uncertainties in the upper plots of Fig.~\ref{fig:perturbative_sm_wcs}. For completeness Fig.~\ref{fig:perturbative_sm_wcs} also shows resonance contributions as in Eq.~\eqref{eq:resonances} in the lower plot. $C^{\text{eff}}_9$ only varies significantly for $q^2$ values below $0.1\,\text{GeV}^2$. For $q^2>0.2$ both real and imaginary parts of $C^{\text{eff}}_9$ remain small and constant at the percent level. $C^{\text{eff}}_7$ varies in the full kinematic region, but the overall magnitude is permille or less. Resonance contributions peak at $q^2=m_\phi$ and $q^2=m_{\rho,\,\omega}$ as expected, where values of order 10 are possible. Also away from the resonances, $C_9^R$ can be an order one contribution, such that contributions from $C^{\text{eff}}_7$ and $C^{\text{eff}}_9$ are negligible, except for $F_L$ where uncertainties from $C_9^R$ contributions cancel and interference effects of $C^{\text{eff}}_7$ and $C_9^R$ give the dominant uncertainties in the SM, see Fig.~\ref{fig:flsm}.

\section{Form Factors}\label{app:FF}

\begin{figure}[!t]\centering
\includegraphics[width=0.49\textwidth]{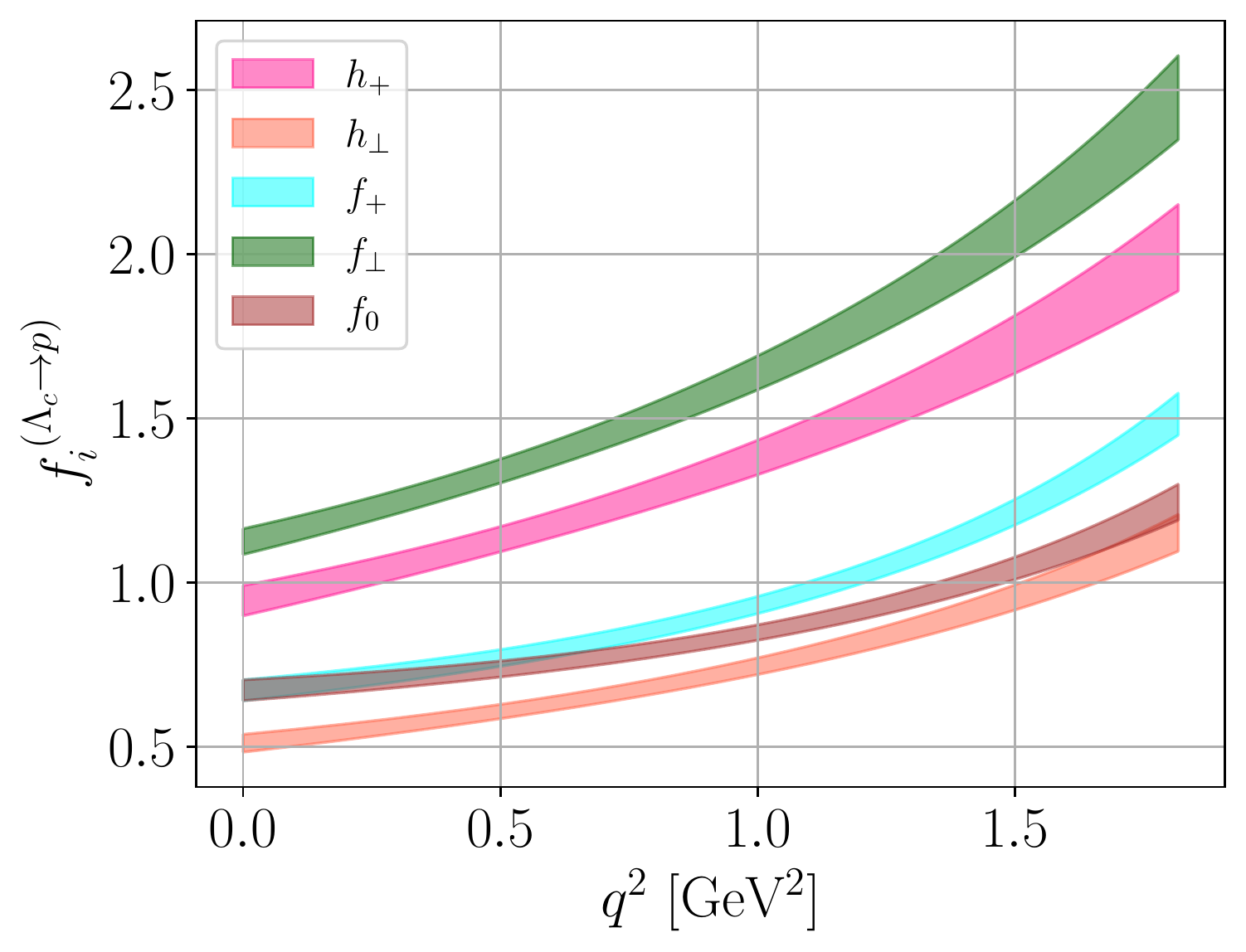}
\includegraphics[width=0.49\textwidth]{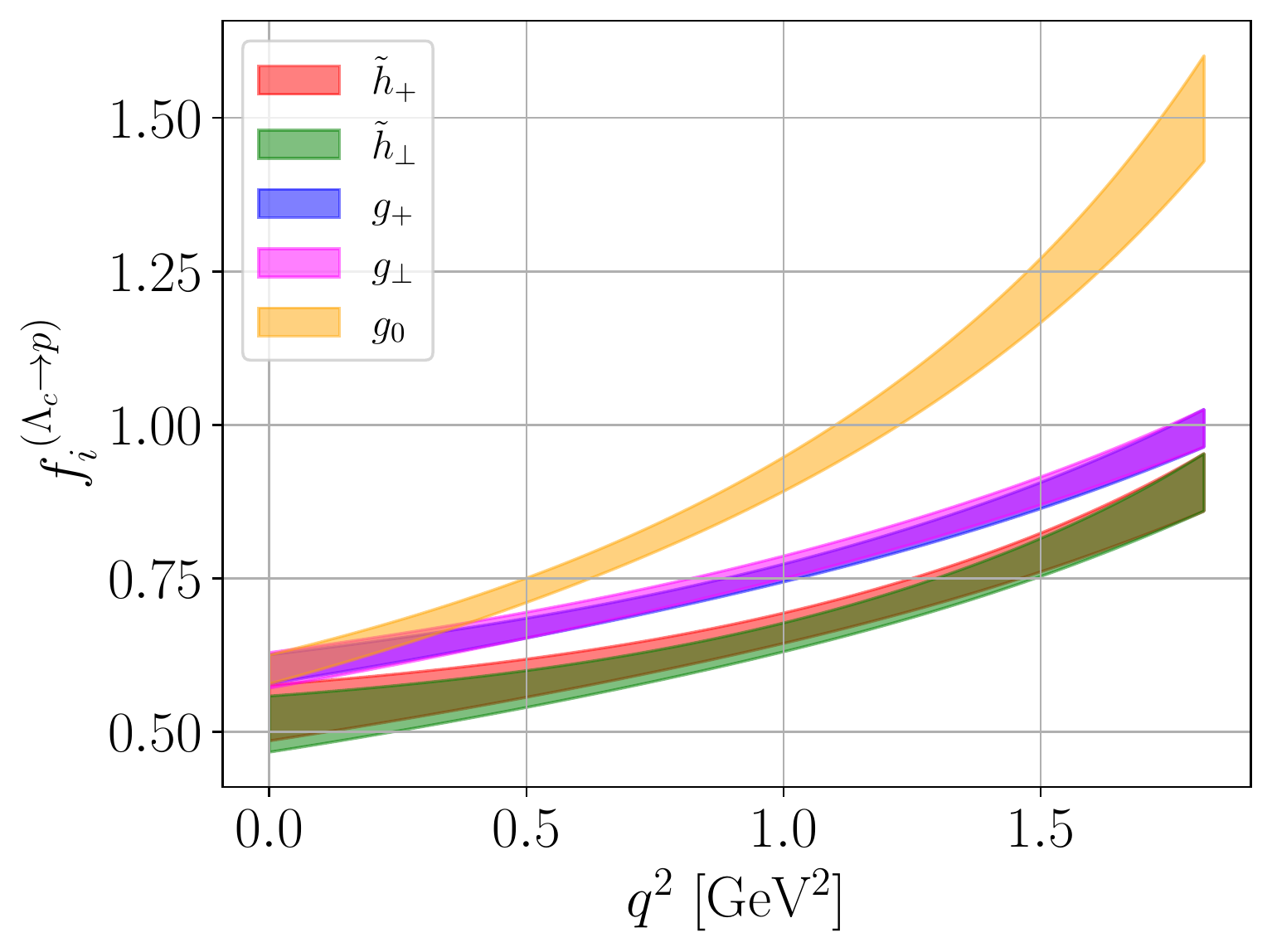}
\caption{The $\Lambda_c\to p$ form factors from lattice QCD~\cite{Meinel:2017ggx} with  $1\sigma$ uncertainties.}
\label{fig:FF}
\end{figure}

The form factors for $\Lambda_c \to p$ in the helicity-based definition read 
\begin{align}
\begin{split}
\langle p(k,\,s_{p}) \vert \overline{u} \gamma^\mu c\vert \Lambda_c(p,\,s_{\Lambda_c}) \rangle = &\\
\overline{u}_{p}(k,\,s_{p})&\left[f_0(q^2)\,(m_{\Lambda_c}-m_{p})\,\frac{q^\mu}{q^2}\right.\\
&\left. + f_+(q^2)\,\frac{m_{\Lambda_c}+m_{p}}{s_+}\,\left(p^\mu + k^\mu - (m_{\Lambda_c}-m_{p})\frac{q^\mu}{q^2}\right)\right.\\
&\left.+ f_{\perp}(q^2)\,\left(\gamma^\mu - \frac{2m_{p}}{s_+}\,p^\mu-\frac{2m_{\Lambda_c}}{s_+}\,k^\mu\right)\right] u_{\Lambda_c}(p,\,s_{\Lambda_c})\,,
\end{split}
\label{eq:ff1}
\end{align}

\begin{align}
\begin{split}
\langle p(k,\,s_{p}) \vert \overline{u} \gamma^\mu\gamma_5 c\vert \Lambda_c(p,\,s_{\Lambda_c}) \rangle=&\\
-\overline{u}_{p}(k,\,s_{p})\gamma_5&\left[g_0(q^2)\,(m_{\Lambda_c}+m_{p})\,\frac{q^\mu}{q^2}\right.\\
&\left. + g_+(q^2)\,\frac{m_{\Lambda_c}-m_{p}}{s_-}\,\left(p^\mu + k^\mu - (m_{\Lambda_c}-m_{p})\frac{q^\mu}{q^2}\right)\right.\\
&\left.+ g_{\perp}(q^2)\,\left(\gamma^\mu + \frac{2m_{p}}{s_-}\,p^\mu-\frac{2m_{\Lambda_c}}{s_-}\,k^\mu\right)\right] u_{\Lambda_c}(p,\,s_{\Lambda_c})\,,
\end{split}
\label{eq:ff2}
\end{align}

\begin{align}
\begin{split}
\langle p(k,\,s_{p}) \vert \overline{u} \text{i} \sigma^{\mu\nu}q_\nu c\vert \Lambda_c(p,\,s_{\Lambda_c}) \rangle =&\\
-\overline{u}_{p}(k,\,s_{p})& \left[h_+(q^2)\frac{q^2}{s_+}\left(p^\mu+k^\mu-(m_{\Lambda_c}^2-m_{p}^2)\frac{q^\mu}{q^2}\right)\right.\\
&\left.+h_\perp(q^2)(m_{\Lambda_c}+m_{p})\left(\gamma^\mu-\frac{2m_{p}}{s_+}\,p^\mu-\frac{2m_{\Lambda_c}}{s_+}\,k^\mu\right)\right] u_{\Lambda_c}(p,\,s_{\Lambda_c})\,,
\end{split}
\label{eq:ff3}
\end{align}

\begin{align}
\begin{split}
\langle p(k,\,s_{p}) \vert \overline{u} \text{i} \sigma^{\mu\nu}q_\nu \gamma_5 c\vert \Lambda_c(p,\,s_{\Lambda_c}) \rangle =&\\
-\overline{u}_{p}(k,\,s_{p})\gamma_5& \left[\tilde{h}_+(q^2)\frac{q^2}{s_-}\left(p^\mu+k^\mu-(m_{\Lambda_c}^2-m_{p}^2)\frac{q^\mu}{q^2}\right)\right.\\
&\left.+ \tilde{h}_\perp(q^2)(m_{\Lambda_c}-m_{p})\left(\gamma^\mu+\frac{2m_{p}}{s_-}\,p^\mu-\frac{2m_{\Lambda_c}}{s_-}\,k^\mu\right)\right] u_{\Lambda_c}(p,\,s_{\Lambda_c})\,,
\end{split}
\label{eq:ff4}
\end{align}
where $q^\mu=p^\mu-k^\mu$. 
The form factors obey the following endpoint relations
\begin{align}\label{eq:endpoint_rel}
\begin{split}
f_0(0)&=f_+(0)\,,\\
g_0(0)&=g_+(0)\,,\\
h_\perp(0)&=\tilde h_\perp(0)\,,\\
g_\perp(q_{\text{max}}^2)&=g_+(q_{\text{max}}^2)\,,\\
\tilde{h}_\perp(q_{\text{max}}^2)&=\tilde{h}_+(q_{\text{max}}^2)\,.
\end{split}
\end{align}
The relation between the dipole form factors $h_\perp, \tilde h_\perp$ at $q^2=0$ follows from
 $\sigma^{\mu \nu} \gamma_5=-i/2 \epsilon^{\mu \nu \alpha \beta} \sigma_{\alpha \beta}$, and is consistent with \cite{Gutsche:2013pp,HillerZwicky21} using
 other form factor parameterizations, however,  is missing in \cite{Meinel:2017ggx}.  Since in the latter works the relation is numerically satisfied within uncertainties $ h_\perp(0) =0.511\pm0.027,\, \tilde h_\perp(0)=0.51\pm0.05$, and, as we keep finite lepton masses and never reach $q^2=0$ exactly anyway,  this has negligible effect on our analysis.
In Ref.~\cite{Meinel:2017ggx} and supplemented files, fit results and correlation matrices are given  in terms of the $z$-expansion
\begin{align}
f(q^2)=\frac{1}{1-q^2/(m^f_{\text{pole}})^2}\,\sum_{n=0}^2\,a_n^f\,[z(q^2)]^n,
\end{align}
where
\begin{align}\label{eq:zexpansion}
\begin{split}
z(q^2) &= {\sqrt{t_+-q^2} - \sqrt{t_+-t_0} \over \sqrt{t_+-q^2} + \sqrt{t_+-t_0}}\,, 
\end{split}
\end{align}
$t_+ =(m_D + m_\pi)^2$ and $t_0 = (m_{\Lambda_c}-m_p)^2$. The value of $m_{\text{pole}}^f$ needs to be picked differently for each form factor, see Table IV in Ref.~\cite{Meinel:2017ggx}. Resulting form factors are shown in Fig.~\ref{fig:FF}.
Form factor relations \eqref{eq:endpoint_rel} hold for $f_0$ and $f_+$ (left plot, cyan and brown) and for $g_0$ and $g_+$ (right plot, yellow and blue), 
and at the zero recoil endpoint $g_\perp=g_+$ (right plot, blue and magenta) and $\tilde{h}_\perp=\tilde{h}_+$ (right plot, red and green).

\section{Helicity Amplitudes}\label{app:hel_amp}

The angular observables in Eq.~\eqref{eq:angl_distr} can be expressed in terms of helicity amplitudes $\mathcal{H}^{m}_{\lambda_p,\lambda_\gamma}$, where $\lambda_p$ and $\lambda_\gamma$ denote the helicities of the proton $p$ and the effective current $\gamma^* (\to \ell\ell)$, respectively. $\lambda_p$ can therefore take values of $\pm\frac{1}{2}$, whereas $\lambda_\gamma$ takes values of $0,\,\pm1$ and we further distinguish $\lambda_\gamma = t$ in the $J_\gamma = 0$ case and  $\lambda_\gamma = 0$ in the $J_\gamma = 1$ case. The index $m$ distinguishes between leptonic vector ($m=1$) and axial vector ($m=2$) contributions. The $m = 2$ amplitudes thus only contain $C_{10}$ and $C_{10}^\prime$.

Following Ref.~\cite{Gutsche:2013pp} we introduce ($m^{(\prime)}= 1,\,2$)
\begin{align}
  \label{eq:1}
  \begin{split}
    &S^{m m'} = N^2\cdot \text{Re}\bigg[\mathcal{H}^{m}_{\frac{1}{2},\,t}\mathcal{H}^{\dagger m'}_{\frac{1}{2},\,t} + \mathcal{H}^{m}_{-\frac{1}{2},\,t}\mathcal{H}^{\dagger m'}_{-\frac{1}{2},\,t}\bigg],\\
    &U^{m m'} =N^2\cdot \text{Re}\bigg[\mathcal{H}^{m}_{\frac{1}{2},\,1}\mathcal{H}^{\dagger m'}_{\frac{1}{2},\,1} + \mathcal{H}^{m}_{-\frac{1}{2},\,-1}\mathcal{H}^{\dagger m'}_{-\frac{1}{2},\,-1}\bigg],\\
    &P^{m m'} =N^2\cdot \text{Re}\bigg[\mathcal{H}^{m}_{\frac{1}{2},\,1}\mathcal{H}^{\dagger m'}_{\frac{1}{2},\,1} - \mathcal{H}^{m}_{-\frac{1}{2},\,-1}\mathcal{H}^{\dagger m'}_{-\frac{1}{2},\,-1}\bigg],\\
    &L^{m m'} =N^2\cdot \text{Re}\bigg[\mathcal{H}^{m}_{\frac{1}{2},\,0}\mathcal{H}^{\dagger m'}_{\frac{1}{2},\,0} + \mathcal{H}^{m}_{-\frac{1}{2},\,0}\mathcal{H}^{\dagger m'}_{-\frac{1}{2},\,0}\bigg],
  \end{split}
\end{align}

To obtain the  helicity amplitudes we sum the contributions from different hadronic matrix elements. To do so we define amplitudes corresponding to one single hadronic matrix element as $\mathcal{H}^{a,m}_{\lambda_p, \lambda_\gamma}$, and $\mathcal{H}^{m}_{\lambda_p, \lambda_\gamma}=\sum_a \mathcal{H}^{a,m}_{\lambda_p, \lambda_\gamma}$. 
$a=1\,(2)$ corresponds to matrix elements involving $C_7^{(\prime)}$ with (without $\gamma_5$). The same notation distinguishes vector and axial vector matrix elements of $C_{9\,(10)}^{(\prime)}$ for $a=3,\,4$. Flipping the helicities will result in a minus sign for the amplitudes $a=2$ and $a=4$, due to parity
\begin{align}
\begin{split}
  &\mathcal{H}^{1}_{\lambda_p,\lambda_\gamma} = \mathcal{H}^{{1},{1}}_{\lambda_p,\lambda_\gamma} + \mathcal{H}^{{2},{1}}_{\lambda_p,\lambda_\gamma} + \mathcal{H}^{{3},{1}}_{\lambda_p,\lambda_\gamma} + \mathcal{H}^{{4},{1}}_{\lambda_p,\lambda_\gamma},\\
  &\mathcal{H}^{1}_{-\lambda_p,-\lambda_\gamma} = \mathcal{H}^{{1},{1}}_{\lambda_p,\lambda_\gamma} - \mathcal{H}^{{2},{1}}_{\lambda_p,\lambda_\gamma} + \mathcal{H}^{{3},{1}}_{\lambda_p,\lambda_\gamma} - \mathcal{H}^{{4},{1}}_{\lambda_p,\lambda_\gamma},\\
  &\mathcal{H}^{2}_{\lambda_p,\lambda_\gamma} = \mathcal{H}^{{3},{2}}_{\lambda_p,\lambda_\gamma} + \mathcal{H}^{{4},{2}}_{\lambda_p,\lambda_\gamma},\\
  &\mathcal{H}^{2}_{-\lambda_p,-\lambda_\gamma} = \mathcal{H}^{{3},{2}}_{\lambda_p,\lambda_\gamma} - \mathcal{H}^{{4},{2}}_{\lambda_p,\lambda_\gamma}\,.
\end{split}
\end{align}

\noindent For convenience we give the list of single contributions below.

\noindent
{$\lambda_{\Lambda_c} = \frac{1}{2}$, $\lambda_{\gamma} = t$:}
\begin{align}
  \begin{split}
  &\mathcal{H}^{1,1}_{-\frac{1}{2},t} = 0\,,\\
  &\mathcal{H}^{2,1}_{-\frac{1}{2},t} = 0\,,\\
  &\mathcal{H}^{3,1(2)}_{-\frac{1}{2},t} = (C_{9(10)}+C_{9'(10')})\frac{\sqrt{s_+}}{\sqrt{q^2}}f_0(q^2)(m_{\Lambda_c}-m_p)\,,\\
  &\mathcal{H}^{4,1(2)}_{-\frac{1}{2},t} = (C_{9(10)}-C_{9'(10')})\frac{\sqrt{s_-}}{\sqrt{q^2}}g_0(q^2)(m_{\Lambda_c}+m_p)\,,
  \end{split}
\end{align}
{$\lambda_{\Lambda_c} = -\frac{1}{2}$, $\lambda_{\gamma} = 0$:}
\begin{align}
  \begin{split}
  &\mathcal{H}^{1,1}_{\frac{1}{2},0} = (C_7+C_7^\prime)\frac{2m_c}{\sqrt{q^2}}\sqrt{s_-}h_+(q^2)\,,\\
  &\mathcal{H}^{2,1}_{\frac{1}{2},0} = -(C_7-C_7^\prime)\frac{2m_c}{\sqrt{q^2}}\sqrt{s_+}\tilde{h}_+(q^2)\,,\\
  &\mathcal{H}^{3,1(2)}_{\frac{1}{2},0} = (C_{9(10)}+C_{9'(10')})\frac{1}{\sqrt{q^2}}\sqrt{s_-}f_+(q^2)(m_{\Lambda_c}+m_p)\,,\\
  &\mathcal{H}^{4,1(2)}_{\frac{1}{2},0} = -(C_{9(10)}-C_{9'(10')})\frac{1}{\sqrt{q^2}}\sqrt{s_+}g_+(q^2)(m_{\Lambda_c}-m_p)\,,
  \end{split}
\end{align}
{$\lambda_{\Lambda_c} = \frac{1}{2}$, $\lambda_{\gamma} = 1$:}
\begin{align}
  \begin{split}
  &\mathcal{H}^{1,1}_{\frac{1}{2},1} = \sqrt{2}(C_7+C_7^\prime)\frac{2m_c}{q^2}\sqrt{s_-}h_\bot(q^2)(m_{\Lambda_c}+m_p)\,,\\
  &\mathcal{H}^{2,1}_{\frac{1}{2},1} = -\sqrt{2}(C_7-C_7^\prime)\frac{2m_c}{q^2}\sqrt{s_+}\tilde{h}_\bot(q^2)(m_{\Lambda_c}-m_p)\,,\\
  &\mathcal{H}^{3,1(2)}_{\frac{1}{2},1} = \sqrt{2}(C_{9(10)}+C_{9'(10')})\sqrt{s_-}f_\bot(q^2)\,,\\
  &\mathcal{H}^{4,1(2)}_{\frac{1}{2},1} = -\sqrt{2}(C_{9(10)}-C_{9'(10')})\sqrt{s_+}g_\bot(q^2)\,.
  \end{split}
\end{align}

\section{Lepton flavor violating angular observables}\label{app:lfv}

For the lepton flavor violating decays the angular observables in equation~\eqref{eq:angl_distr} are given by
\begin{align}
  \label{eq:4}
  \begin{split}
    K_{1ss} = ~ &2(m_{\ell} - m_{\ell'})^{2}v^{2}_{+}S^{11} + 2(m_{\ell} + m_{\ell'})^{2}v^{2}_{-}S^{22}\\
    &+ 2q^{2}v^{2}_{-}L^{11} + 2q^{2}v^{2}_{+}L^{22}\\
    &+ \left(q^{2}v^{2}_{-} + (m_{\ell} + m_{\ell'})^{2}v^{2}_{-}\right)U^{11}
    + \left(q^{2}v^{2}_{+} + (m_{\ell} - m_{\ell'})^{2}v^{2}_{+}\right)U^{22}, \\
    K_{1cc} = ~ &2(m_{\ell} - m_{\ell'})^{2}v^{2}_{+}S^{11} + 2(m_{\ell} + m_{\ell'})^{2}v^{2}_{-}S^{22} \\
    &+ 2(m_{\ell} + m_{\ell'})^{2}v^{2}_{-}L^{11} + 2(m_{\ell} - m_{\ell})^{2}v^{2}_{+}L^{22}\\
    &+ 2q^{2}v^{2}_{-}U^{11} + 2q^{2}v^{2}_{+}U^{22}, \\
    K_{1c} = ~&-4q^{2}v_{+}v_{-}P^{12},
  \end{split}
\end{align}
where we use the expressions from Eq.~\eqref{eq:1} for the hadronic helicity amplitudes with the appropriate replacement of Wilson coefficients $C_{9,\,10}^{(\prime)}$ with lepton flavor violating Wilson coefficients $K_{9,\,10}^{(\prime)}$, skipping contributions with $C_7^{(\prime)}$ and using $v_{\pm}~=~\sqrt{1 - \frac{(m_{\ell} \pm m_{\ell'})^{2}}{q^{2}}}$.

\end{document}